\documentclass[11pt,twocolumn]{article}
\usepackage[utf8]{inputenc}

\usepackage{a4}
\usepackage{amsmath}
\usepackage{graphicx}
\usepackage{mathtools}
\usepackage{epstopdf}
\usepackage{cuted}
\usepackage{float}
\usepackage{hyperref}
\usepackage{abstract}
\usepackage{subcaption}
\usepackage[toc,page]{appendix}
\usepackage[font=it]{caption}
\usepackage[margin=1.0in]{geometry}
\usepackage{verbatim}

\newcommand{\unit}[1]{\ensuremath{\, \mathrm{#1}}}

\title{\bf Simulating Tidal Interactions\\ between Galaxies:\\ A Pre-University Student Project}

\author{
    M. Brea-Carreras, M. Thiel \& M. P\"ossel\footnote{Corresponding author: poessel@hda-hd.de}}

\date{\small Haus der Astronomie, MPIA-Campus, K\"onigstuhl 17, 69117 Heidelberg, Germany	}

\begin{document}

\maketitle
 \begin{strip}
\begin{abstract}
We report on a project undertaken in Summer 2017 by pre-university student interns at Haus der Astronomie: point-particle simulations of galaxy collisions with the aim of reproducing observational data from such collisions. We succeeded in providing a visually similar representation of both NGC 5426/7 and the "Antennae" galaxies (NGC 4038/9), and were able to make deductions about the relative positions and orbits of these galaxies. The project is an example for how participants with little more than high-school level previous knowledge can successfully tackle, and understand, advanced topics from current astrophysical research. This report was written by the two participants (M. B.-C. and M. T.), on whose experiences it is based, in collaboration with their supervisor at Haus der Astronomie (M. P.).
    \end{abstract}
\end{strip}

\section{Introduction} \label{introduction}
There is a broad consensus among science educators about the importance for science education of hands-on activities, elements of inquiry-based learning, and understanding science as a process \cite{Beyond2000,Rocard2007,NRC2011}. For high-school students, or students in limbo between high school and university, internships at research institutions are a suitable way of experiencing first hand the process of science, whether as part of a larger project or in the course of working on a dedicated student research project.

Haus der Astronomie has offered research internships since 2009, and is currently offering a yearly ``International Astronomy Summer Internship Program'' for advanced high school students, or to students transitioning from high school to college/university.\footnote{Up-to-date information can be found on 
\href{http://www.haus-der-astronomie.de/en/what-we-do/internships/summer-internship}{http://www.haus-der-astronomie.de/en/what-we-do/internships/summer-internship} } Among the authors of this e-print, two of us attended the 2017 internship program (M. B.-C. and M. T.), while one of us was the internship supervisor (M. P.).\footnote{During the first three weeks of the 2017 internship, the project was worked on by M. T. and another intern; during the second three weeks, M. T. and M. B.-C. continued the project in their spare time, on their own initiative. This unusual, and remarkably successful, extra effort prompted M. P. to have the two interns present their results at the WE Heraeus Summer School ``Astronomy From Four Perspectives: The Dark Universe'' in Heidelberg. In summer 2018, this project was also presented by M. T. at the German national teacher training in astronomy at the University of Jena.} 

The project described here used computer simulations to reproduce classic results about galaxy evolution and the observational properties of interacting galaxies. Pedagogically, numerical treatments have several advantages \cite{FeynmanLectures,1978AmJPh..46..748C,1984AmJPh..52..499S}: They allow students to explore dynamical situations that are much too advanced for a direct solution of the differential equations involved. In fact, when it comes to the underlying mathematics, the approach is easier to understand than most techniques for dealing with differential equations analytically. Numerical approaches also allow us to describe situations that are beyond the reach of even the most advanced methods for finding analytical solutions (that is, finding solutions that can be written down in terms of basic, well-known functions). The associated challenge is, of course, that in order to write such simulations themselves (as opposed to using existing software), students will need to have appropriate programming skills.

This article serves a double function: For one, it describes the student research project undertaken by M. B.-C. and M. T., to whom the collective ``we'' in the following sections \ref{SummarySection} to \ref{EndProject} refers; this aspect, we hope, will make the text of interest to those involved in astronomy education,
who might be interested in including this or similar student activities in their own programs. In addition, the e-print is part of the project itself. After all, writing up your methods and results is an integral part of scientific work --- even if, in this case, we have included additional aspects related to astronomy education which are not found in ordinary astronomy papers. Each intern presents a personal view of their experiences in the appendix, sections \ref{Brea} and \ref{Thiel}. 

The scripts, simulations parameters and presentations written for this project are publicly available and can be downloaded from GitHub\footnote{The repository can be cloned from
\href{https://github.com/mbrea-c/simulating-tidal-interactions.git}{https://github.com/mbrea-c/simulating-tidal-interactions.git} }.

\section{Project summary}
\label{SummarySection}

Theories of galaxy evolution and formation need to be able to explain, among other things, the appearance of interacting galaxies. The most widely accepted explanation was first studied more intensively in the 1960s \cite{pfleiderer1960spiral}, and identifies tidal interactions between galaxies as the cause of the deformations of the galaxies involved. The main reference for our student research project is a classic 1972 paper by Alar and Juri Toomre  \cite{toomre1972galactic}, which describes basic numerical simulations of galaxy interactions. The authors argue that the bridges and tails visible in some spiral galaxies are remnants of tidal interactions between galaxies involved in a close encounter --- and they support their argument with restricted three-body calculations of selected encounters, including reconstructions of a few well-known pairs: the ``Antennae'' galaxies NGC4039/39, the ``Mice'' galaxies NGC 4676, and M51 with its small companion NGC 5195. 

For our research project, we followed in the footsteps of Toomre and Toomre by creating our own galaxy simulation, written using the Python programming language.\footnote{Python Software Foundation. Python Language Reference, version 2.7. Available at \href{http://www.python.org}{http://www.python.org}} We describe our simulation techniques in section \ref{methods}, while section \ref{simulations} presents our results. In section \ref{education}, we draw on our personal experience to give recommendations for similar educational projects. 

 \section{Methods} \label{methods}

Our simulations were written in Python. We are aware that using a general programming language like Python as the basis of a student research project raises the bar higher than relying on more specialized and restricted application software, specifically for students with no previous programming experience. In our case, one of us (M. B.-C.) had previous experience in using Python, while the other (M. T.) learned Python during his internship, but had some previous experience programming in Java. Going by our personal experience, we think that Python is a good choice for those who are new to programming.

The simulations were all three-dimensional, in the sense that all vector data stored and used in the computations was three-dimensional. However, in the first two examples all particles and velocities are confined to the same plane, making those simulations two-dimensional in effect.

The basic elements of our physical model are point particles with mass on the one hand and point particles without mass on the other hand. The latter serve as test particles, which are acted upon by gravitational forces but are not themselves treated as sources of gravity. Following Toomre and Toomre, the mass of each galaxy is represented by a single point particle with the total galaxy mass, which is located in the center of the galaxy. Galaxy structure is simulated using test particles representing stars in each galaxy's (initially undisturbed) disk.

It is clear that this will capture only part of what happens when two galaxies collide -- interactions between the outer regions of the two galaxies are not part of this simplified scenario. Still, as Toomre and Toomre found and as we reproduce with our simulations, even this simplified model can reproduce some of the features of colliding galaxies that are found in actual observations.
        
With this simplification, there are just two sources of gravity in a two-galaxy simulation: the two central point particles carrying each galaxy's mass. For all other particles, we only need to calculate how they react to the gravitational forces exerted by the two central point particles. 
 
       For each particle, we need to store its three-dimensional position and its velocity. In the case of the central particles of each galaxy, their is an addtional parameter: the scalar mass. This data, along with a predefined time-step, is passed to an update function, which returns the particles' new positions and velocities. Results are plotted using the Python 2D plotting library matplotlib \cite{Hunter:2007}, which we found simple and straightforward to use for our purposes.
     
        \subsection{Update Function}
        
        In order to define a physical update function for a simulation -- which encapsulates how our system changes over time -- we need to perform a second order integration: to solve for the trajectory of moving bodies affected by a location-dependent acceleration. Whereas the two-body problem for two point particles orbiting each other under the influence of their mutual gravitational attraction can be solved analytically, the general situation of three or more bodies can only be described numerically.
        
In our particular set-up, we have $N$ particles, each designated by an index $j$. We denote the $j$th particle's position at time $t$ as $\vec{x}_j(t)$, its velocity at time $t$ by $\dot{\vec{x}}_j(t)$ and its acceleration by $\ddot{\vec{x}}_j(t)$. 

In general, the acceleration felt by the $j$th particle at time $t$ would result from the sum of all the other particles exerting a gravitational force; each separate gravitational influence is described by Newton's inverse-square law of gravity. The total acceleration of the $j$th particle is 
        \begin{equation}
                \ddot{\vec{x}}_j(t) =- G\sum_{i \neq j} m_i\cdot  \frac{\vec{x}_{j}(t) - \vec{x}_{i}(t)\phantom{{}^3}}{|\vec{x}_{j}(t) - \vec{x}_i(t)|^3},
                \label{GravitationalAcceleration}
        \end{equation}
 where $m_i$ is the mass of the $i$th particle and $G$ denotes the gravitational constant. In our simplified situation, most of the particles are massless, with $m_i=0$, and thus do not act as sources of gravity. Thus, the sum on the right-hand side of equation (\ref{GravitationalAcceleration}) will run over at most two sources, the two ``galaxy centers of mass.''
 
For the purpose of numerical integration, time is divided into discrete time steps. The update function is used to calculate the state of the system after each additional time step, based on the information available at the present time step or earlier time steps. We define the elapsed time at time-step $n$ as
         \begin{equation}
         t_n = t_0 + n h
         \end{equation}
where $h$ denotes the (in our case: universal) time interval between one time step and the next. This discretisation is a simplification, and it stands to 
reason that the value defined for $h$ will have direct consequences for the accuracy of any numerical calculation.

            \subsubsection{Euler method}
            
            	\begin{figure}
                \centering
  				\includegraphics[width=0.5\textwidth]{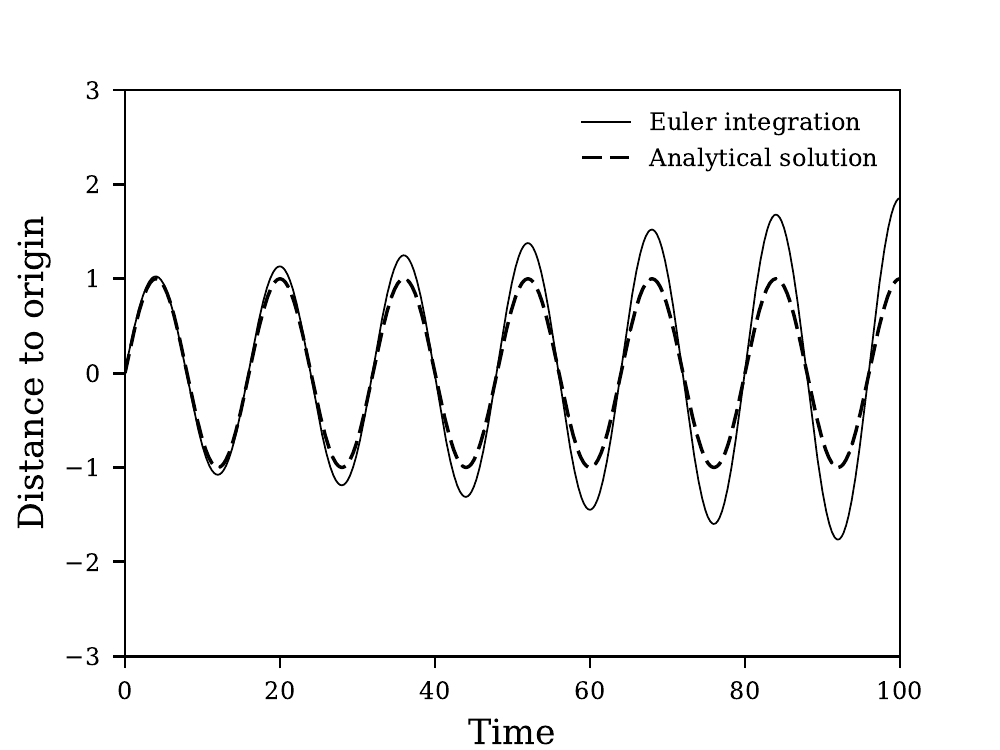}
  				\caption{Displacement from origin over time using Euler integration method (step length $h = 0.08$), compared with the
                analytical solution for a simple harmonic oscillator with amplitude $A = 1$ and angular frequency $\omega = \frac{\pi}{8}$.}
  				\label{fig:euler}
				\end{figure}
                
                The most straightforward way of numerical integration is the Euler method, in which all rates of change are approximated as linear, and the change in a function is taken to be the rate of change times the duration of the time step. Following this prescription, the Euler integration algorithm computes the updated velocity and position for each body $j$ and for each iteration $n$  as           

                \begin{eqnarray}
                \dot{\vec{x}}_j(t_{n+1})& =& \dot{\vec{x}}_j(t_n) + h \ddot{\vec{x}}_j(t_n)
                \label{EulerAcceleration}\\[0.5em]
                \vec{x}_j(t_{n+1})& =& \vec{x}_j(t_n) + h \dot{\vec{x}}_j(t_n),
                \end{eqnarray}
                where $n$ is the number of iterations, and thus $n h$ is the elapsed time. At each step, we need to determine each particle's acceleration vector $\ddot{\vec{x}}_{j}(t_n)$ following equation (\ref{GravitationalAcceleration}).
                
			\subsubsection{Numerical errors and the velocity Verlet Algorithm}
			\label{StabilitySection}
			Euler integration incorporates certain systematic errors, which lead to cumulative divergence between a numerical solution and the true solution. An instructive example is that of the one-dimensional harmonic oscillator, where the acceleration is given by
			\begin{equation}
			\ddot{x} = -\omega^2 x
            \label{Hooke}
			\end{equation}
			(Hooke's law) for some constant angular frequency $\omega$. In this simple case, we can directly write down an analytical solution 
			\begin{equation}
			x(t)  = A\cdot \sin(\omega t),
			\end{equation}
			with $A$ the amplitude of the oscillation. This allows for a direct comparison between the analytical and numerical solution which exposes some of the problems inherent in numerical simulations.			
			Imagine that the moving mass of that oscillator is moving towards positive $x$, so that  we are in an phase where $\dot{x}>0$. 
			The Euler prescription (\ref{EulerAcceleration}) assumes that the acceleration acting on our mass during the time step $t_n$ is equal to the acceleration at the {\em beginning} of that time step. But in reality, we know that the effect of the acceleration will be greater than that. After all, during that time step, the mass is moving towards greater x, and the acceleration pulling it back will increase at greater x values. Evidently, the Euler method systematically underestimates the pull experienced by our mass. Conversely, when our mass is moving back towards the origin, the Euler method will systematically overestimate the pull. Both effects erroneously increase the total energy, and thus the amplitude, of the oscillation: the first since it allows the mass to move further outward than allowed, and the second because it imparts a larger momentum on the mass. The cumulative effect can be seen in Fig. \ref{fig:euler}, where the numerical solution progressively diverges from the analytical solution. 
			
Such integration errors have led to the development of alternative numerical integration schemes. One such scheme is the velocity Verlet algorithm \cite{1967PhRv..159...98V}.
                \begin{figure}
                \centering
  				\includegraphics[width=0.5\textwidth]{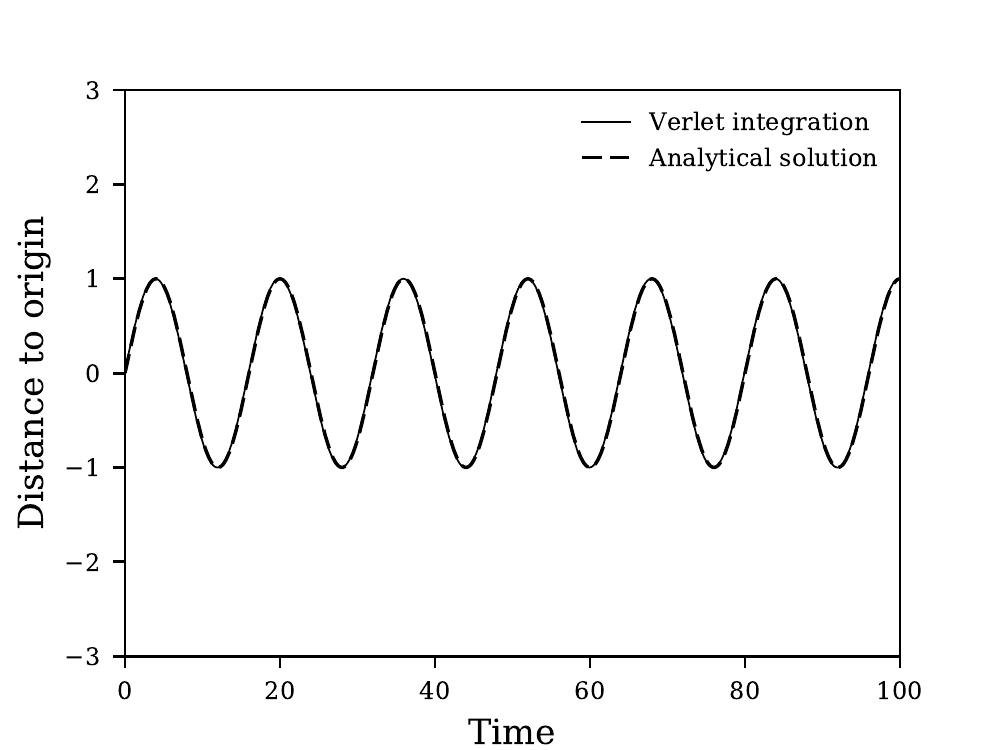}
  				\caption{Displacement to origin over time using Verlet integration method ($h = 0.08$) against the analytical solution of a simple harmonic oscillator $x(t) = A\cdot\sin(\omega t)$, with $A = 1$ and $\omega = \frac{\pi}{8}$).}
  				\label{fig:verlet}
				\end{figure}         
This method introduces additional ''half-steps'' for time, which we will designate by $t_{n+1/2} = (n+1/2)h$. First, the velocity is calculated at one such half-step as
                \begin{equation}
                \dot{\vec{x}}_j(t_{n+1/2}) = \dot{\vec{x}}_j(t_n)  +  \frac{h}{2} \ddot{\vec{x}}_j(t_n)
                \end{equation}
where the acceleration is of course system-specific, eq. (\ref{Hooke}) in the case of the harmonic oscillator, or eq. (\ref{GravitationalAcceleration}) for our galaxy simulation. Once computed, this half-step velocity is used to find the position at $t_{n+1}$ as                 
\begin{equation}
                {\vec{x}}_j(t_{n+1}) = {\vec{x}}_j(t_n) + h\dot{\vec{x}}_j(t_{n+\frac{1}{2}}) 
                \end{equation}
                Next, using these newly-determined position values, the acceleration {\em evaluated at} $t_{n+1}$ is computed, and used to 
                compute  $\dot{\vec{x}}_{n+1}^{(j)}$, as                
                \begin{equation}
				\dot{\vec{x}}_j(t_{n+1}) = \dot{\vec{x}}_j(t_{n + \frac{1}{2}}) + \frac{h}{2} \ddot{\vec{x}}_j(t_{n+1}).
                \end{equation}
For the simple example of the harmonic oscillator, the comparison between Figs. \ref{fig:euler} and \ref{fig:verlet} shows the remarkable increase in numerical stability of this improved algorithm --- at least within the range of the simulation shown in those figures, the Verlet version does not visibly overshoot the mark! The increase in computational cost is a mere 50\%, as the half-steps are used only in the velocity calculation, not in the position calculation.
           
        \subsection{Initial values }
The algorithms described in the previous two sections allow us to simulate the {\em evolution} of physical systems. In order to pick out a particular solution, we need to specify appropriate initial values, as is generally the case when solving differential equations.

The aim of our project is to simulate interactions between galaxies; our initial conditions amount to modelling, in an appropriate way, the original two galaxies involved in the interaction. We make the assumption that, before the interaction proper, we can regard the galaxies as completely separate systems. For each of the galaxies, the test particles are assumed to be in circular orbit around the central galaxy particle. This amounts to another simplification: in real disk galaxies, famously, stellar orbits do {\em not} follow Kepler's law. Instead, rotation speeds become almost constant at greater distances from the galactic center, which is commonly taken as evidence for the existence of Dark Matter in such galaxies.

\subsection{The two-body problem} \label{rmin}

Since we have only the two central particles acting as sources of gravity, it is possible to calculate their motion analytically --- after all, the two-body problem allows for direct analytical solutions. We have made use of this analytical description in choosing the initial conditions for the two central particles, computing the orbital eccentricity $e$ and eventual pericenter distance $R_{min}$ (that is, the distance of closest approach of the galaxy point partices) in each case. 

The key concepts and equations are as follows \cite[chapter 9]{1965mech.book.....K}: Due to the conservation of angular momentum, the two-body motion takes place in the two-dimensional plane defined by the two initial velocity vectors of the central point particles. 
        
For two centers of mass $m_1$ and $m_2$ with positions $\vec{x}_1$ and $\vec{x}_2$, relative position $\vec{r}$ and reduced mass $\mu$ are defined as 
        \begin{equation}
        \vec{r} = \vec{x}_2 - \vec{x}_1
        \end{equation}
        and
        \begin{equation}
        \mu = \frac{m_1 m_2}{m_1 + m_2},
        \end{equation}
        respectively.
Solving the two-body problem amounts to solving two one-body problems: free, linear motion for the center-of-mass of the system, and an effective one-body problem for the relative motion of a particle of mass $\mu$ with position vector $\vec{r}$ in the Newtonian potential of a point mass $M=m_1+m_2$; this latter motion results in a Kepler ellipse. Using polar coordinates $(r,\theta)$, with the origin located at one of the focus points, the shape of the ellipse is described by 
   \begin{equation}
        r = \frac{r_0}{1 - e \cos\theta}.
        \label{EllipseEquation}
        \end{equation}
Here, $\theta=0$ corresponds to the direction from the focus to the center of the ellipse; $e$ is the ellipse's eccentricity.
The quantity $r_0$ is called the {\em semilatus rectum} and sets the length scale of the ellipse.

Each orbit is characterized by two conserved quantities: the mechanical energy
\begin{equation}
        E = \frac{1}{2}\mu \dot{r}^2 - \frac{G m_1 m_2}{r}=\frac{1}{2}\mu \dot{r}^2 - \frac{G \mu M}{r}
        \end{equation}
 and the angular momentum
 \begin{equation}
        L = \mu\cdot \dot{r}_{trans} \cdot r,
        \end{equation}
where $\dot{r}$ is the relative speed of the two centers of mass and $\dot{r}_{trans}$ the component of the relative velocity that is transversal to the location vector.
        
Using $E$ and $L$, the two parameters in equation (\ref{EllipseEquation}) for the shape of the elliptical orbit can be written as
       \begin{eqnarray}
        r_0 &=& \frac{L^2}{\mu G m_1 m_2}\\[0.5em]
        e\phantom{{}_0} &=& \sqrt{1 + \frac{2 E L^2}{\mu (Gm_1m_2)^2}}.
        \end{eqnarray}
Computing the minimal distance $R_{min}$ of our $\mu$ particle from the center of attraction, which is also the minimal distance between our two galaxy particles, amounts to solving an optimization problem, namely
        \begin{equation}
        \frac{\mathrm{d}r}{\mathrm{d}\theta} = \frac{r_0}{1 - e\cos\theta} (e\sin\theta) \stackrel{!}{=}0.
        \end{equation}
The result is      
 \begin{equation}
       \sin\theta = 0.
       \end{equation}
Since $0 \leq \theta < 2\pi$, we must have either $\theta = 0$ or $\theta = \pi$. Inserting these values into (\ref{EllipseEquation}), we find that the minimal value is at $\theta=\pi$, namely
         \begin{equation}
       R_{min} = \frac{r_0}{1 + e}.
       \end{equation}
In our attempts to model certain specific interactive pairs of galaxies, we made use of this relation in choosing (albeit with a certain amount of trial and error) suitable initial values.

\subsection{Optimization}
\label{Optimization}

Even with computers inordinately more powerful than in the early 1970s, when Toomre and Toomre created their simulations, run time values and run time differences in our Python scripts were notable features of our project --- and just like in professional simulation projects, we experimented with ways of optimizing the performance of our programs. 

One area where different implementations turned out to make a considerable difference involved different ways of storing our simulation data. The simplest way of handling simulation data, which we would recommend for students with little to no programming experience, involves the use of Python list objects. The simulation itself involves iterative loops that wrap the update function. 

This implementation turns out to be comparatively slow. In those of our simulations that involved a considerable number of test particles, run time did set the pace both for testing our programs and for finding the best initial values. 

As an alternative, we took advantage of the array structures and associated linear algebra functionality of the NumPy core package within the SciPy ecosystem for Python \cite{SciPyLibrary,NumPy}. User-defined functions such as those necessary to implement our update function can be ''vectorized'' to allow their direct application to array objects, eliminating the need for explicit loops. The linear algebra aspect of that alternative makes it somewhat more challenging for a high school project, but if it is included, the pay-off is a substantial gain in speed, as shown in table \ref{table:extimes}.
            \begin{table}[ht]
            \centering
            \bgroup
            \renewcommand{\arraystretch}{1.2}	
            \begin{tabular}{r | r | r}
            $m$ & Non-vectorized & Vectorized \\ \hline
            28  & 11.291\,s &  1.42\,s \\ 
            134 & 45.404\,s &	 1.511\,s \\
            202 & 67.753\,s &  1.597\,s \\
            268 & 85.136\,s &	 1.621\,s \\
            406 & 126.231\,s & 1.763\,s \\
            540 & 166.262\,s & 1.915\,s \\
            676 & 206.818\,s & 1.993\,s \\
            810 & 251.798\,s & 2.123\,s \\
            \end{tabular}
            \egroup
            \caption{Execution times for increasing particle count $m$ for a constant number of 5000 time steps.
                        \label{table:extimes}}
            \end{table}
In accordance with good computing practices, we also experimented with an object-oriented approach to programming our simulation, as a way of keeping the code modular and easier to organise and debug. In the end though, we found it difficult to reconcile a fully object-oriented approach with the vectorisation of our variables and the update algorithm. 
                    
Last but not least, creating the diagrams used to display the results of our simulations proved a significant contribution to overall run time. We made use of the Matplotlib library \cite{Hunter:2007} for this purpose, which uses the CPU for displaying graphics and thus has a substantial impact on performance. This impact can be reduced when rendering only each $n$th frame, e.g. only every $10^8$ or $10^7$ years of simulated time. An alternative possibility for optimisation, which we did not attempt but which might be of interest to others, would be to switch to 
VPython or PyOpenGL,\footnote{
Information can be found on \href{http://vpython.org/}{http://vpython.org/} and 
\href{http://pyopengl.sourceforge.net/}{http://pyopengl.sourceforge.net/}, respectively.} both of which are GPU-based.

\section{Simulations} \label{simulations}
    
The simulations we created for our project were of two types: in our first step, we set out to replicate two of what Toomre \& Toomre call their  ''elementary examples'' \cite[section II]{toomre1972galactic}, namely the retrograde passage and the direct passage of a small companion. In both cases, there is a main galaxy modelled as a central particle surrounded by a flat disk of orbiting test particles, while the second galaxy is modelled only as a single central particle. In these encounters, the orbital plane of the two central particles coincides with the disk plane of the main galaxy, so the situation as a whole is confined to a two-dimensional plane.

The main part of our project was then dedicated to simulating the interactions in the NGC 5426/7 system and for the "Antennae" (NGC 4038/9) pair of galaxies. In these simulations, each of the interacting galaxies has a disk of orbiting stars, modelled as a disk of test particles, but the simulation is no longer restricted to a two-dimensional plane.
        
        \subsection{Retrograde passage of an equal-mass companion}

            \begin{figure}[!htbp]
            \captionsetup[subfigure]{labelformat=empty}
  			\begin{subfigure}[b]{0.2\textwidth}
    			\includegraphics[width=\textwidth]{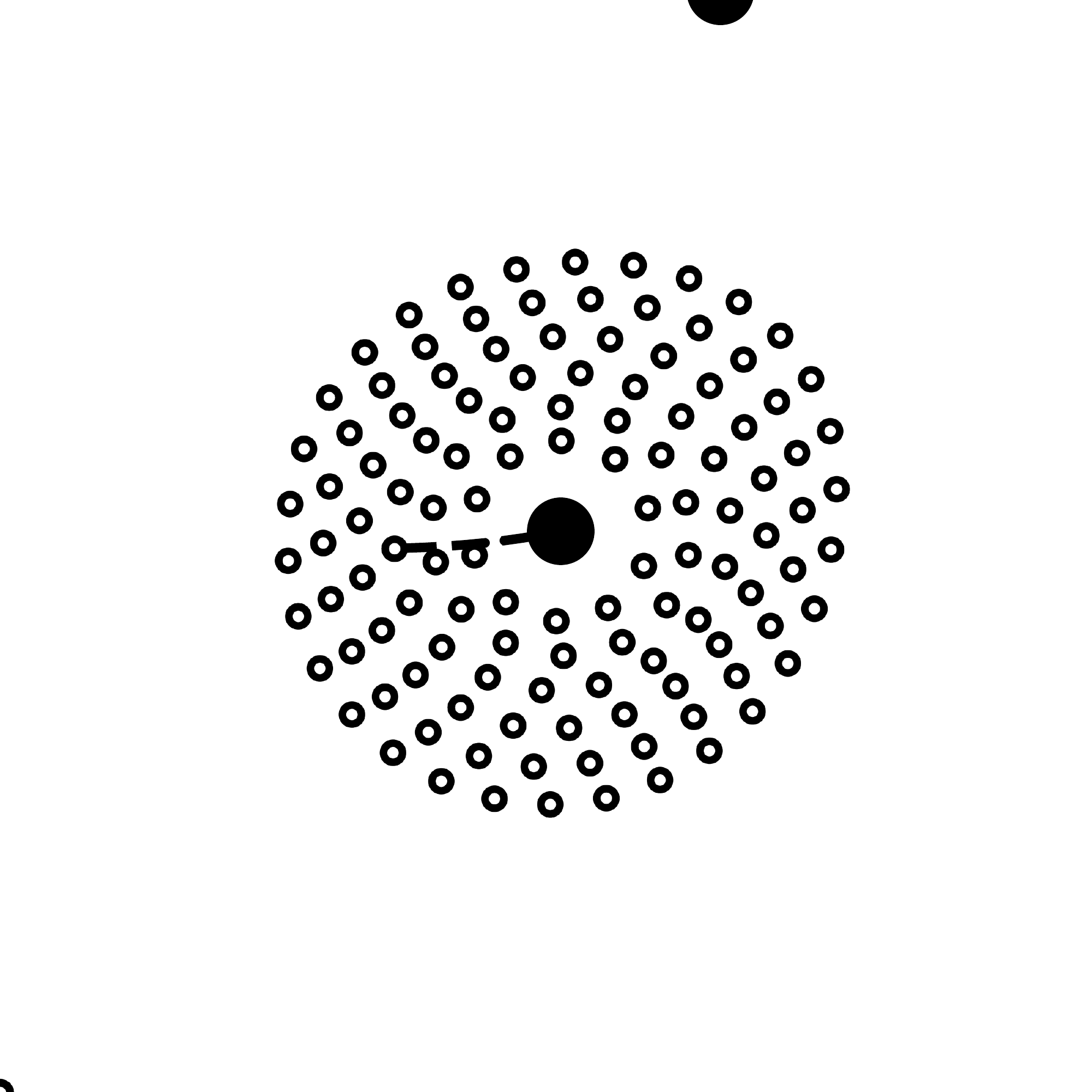}
    			\caption{$t = 1$}
  			\end{subfigure}
  			\hfill
  			\begin{subfigure}[b]{0.2\textwidth}
    			\includegraphics[width=\textwidth]{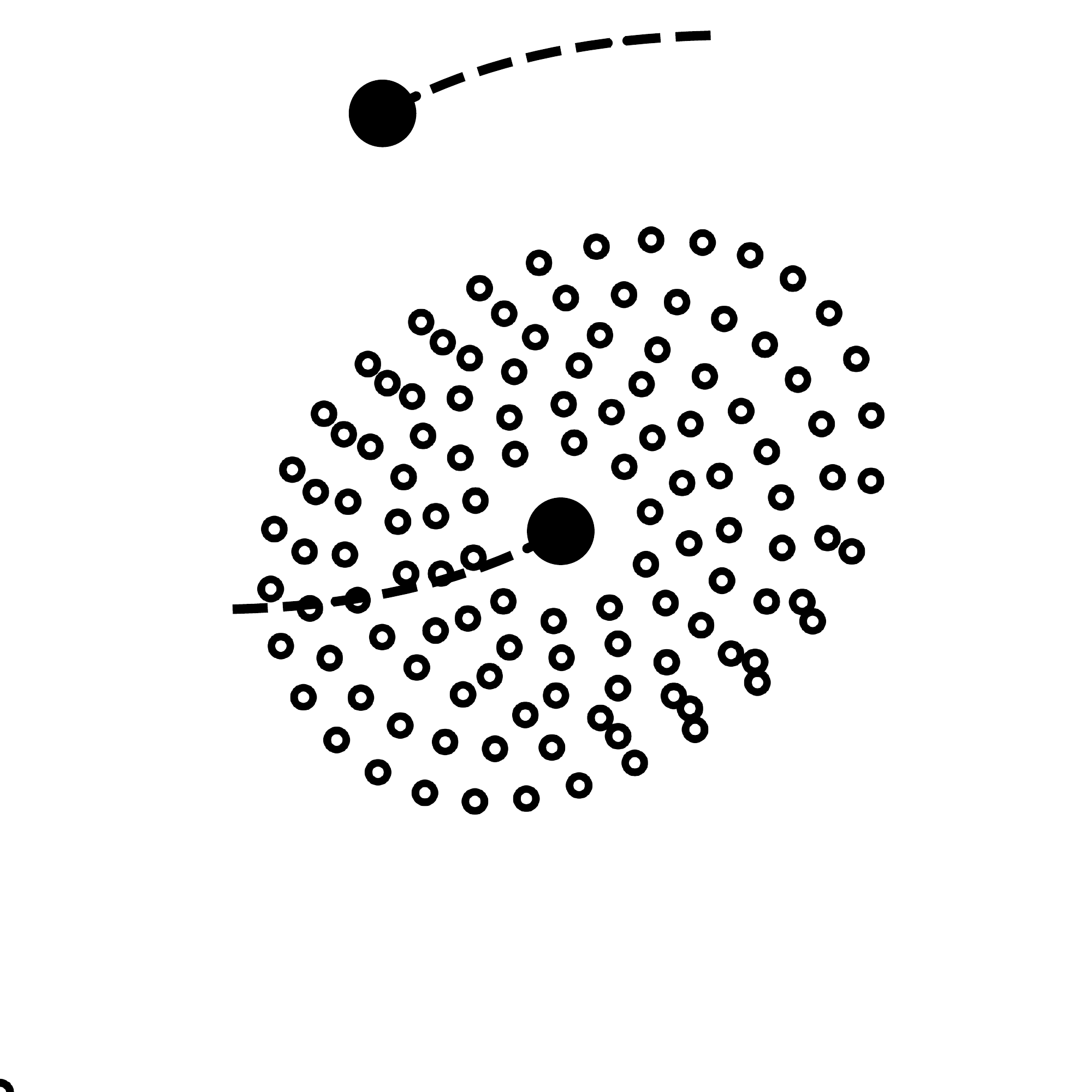}
    				\caption{$t = 2$}
  			\end{subfigure}
            \hfill
            \begin{subfigure}[b]{0.2\textwidth}
    			\includegraphics[width=\textwidth]{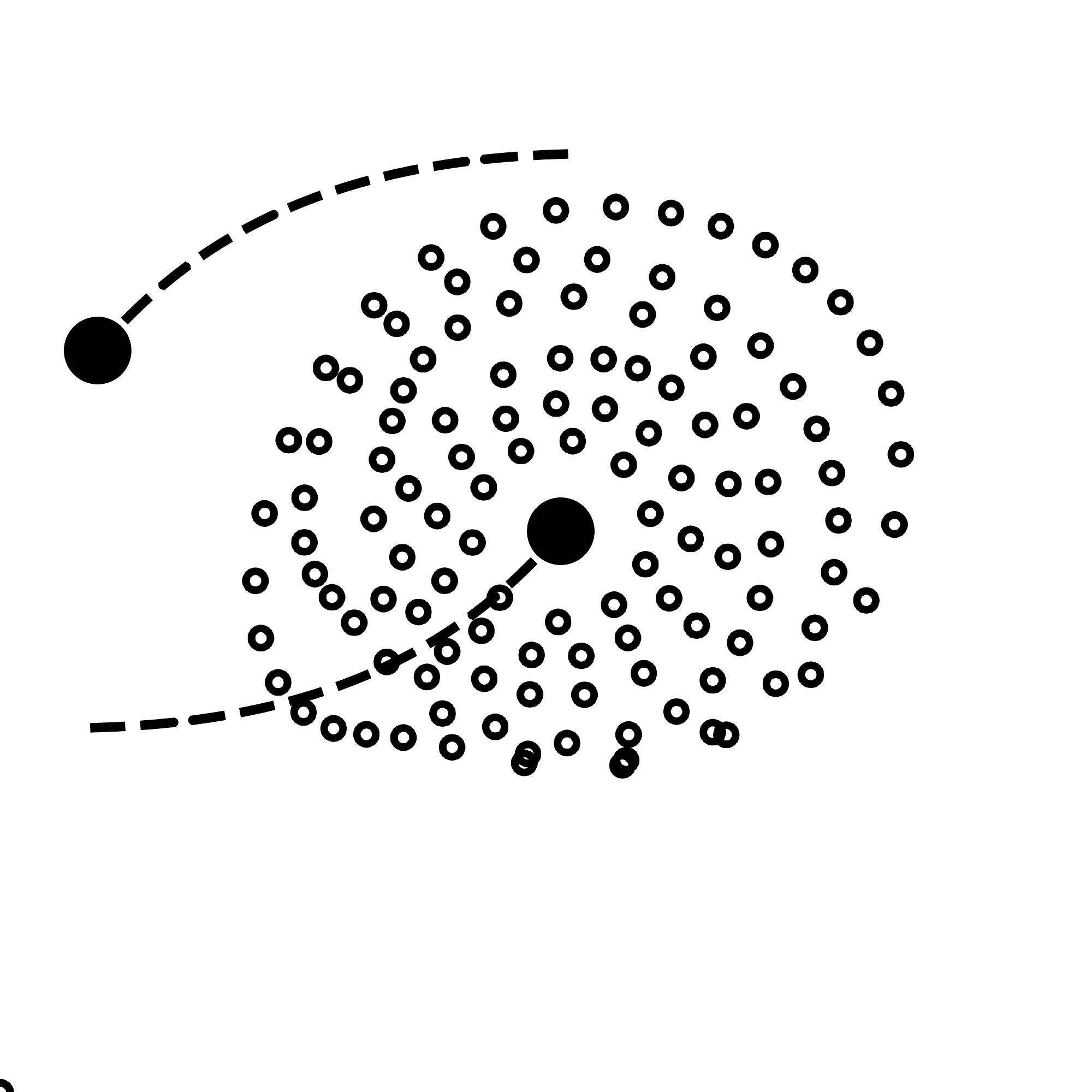}
    				\caption{$t = 3$}
  			\end{subfigure}
            \hfill
            \begin{subfigure}[b]{0.2\textwidth}
    			\includegraphics[width=\textwidth]{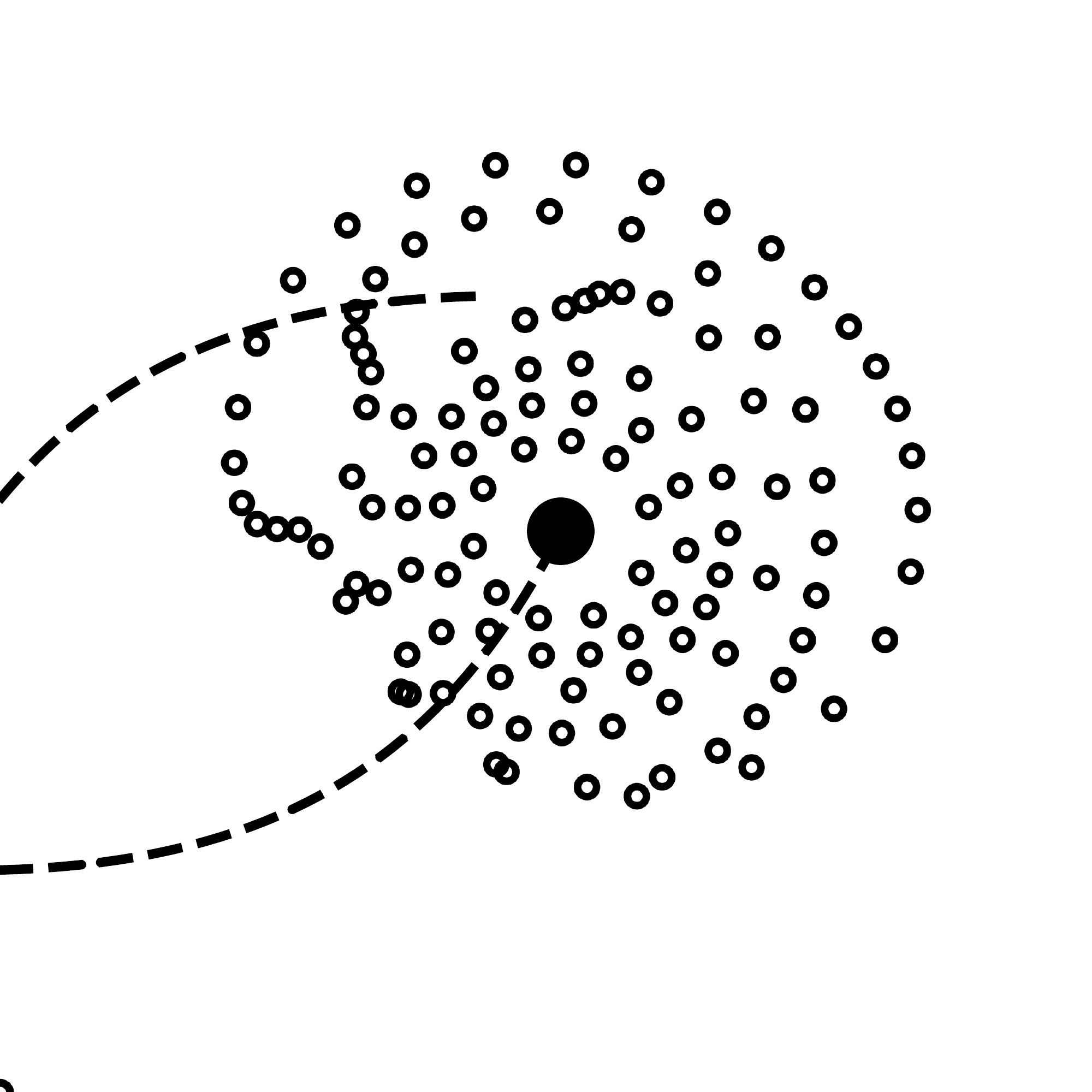}
    				\caption{$t = 4$}
  			\end{subfigure}
  			
            \begin{subfigure}[b]{0.2\textwidth}
    			\includegraphics[width=\textwidth]{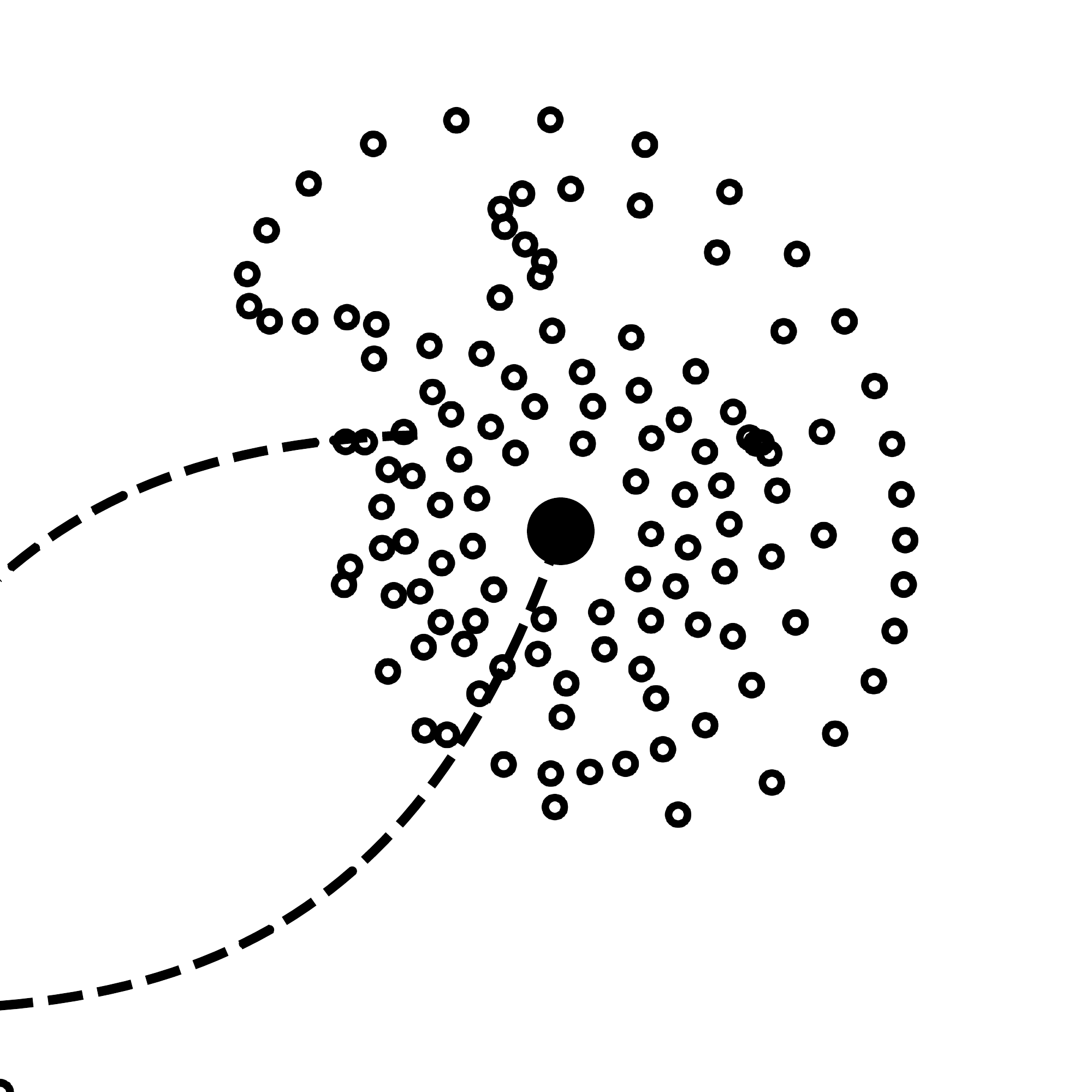}
    				\caption{$t = 5$}
  			\end{subfigure}
            \hfill
            \begin{subfigure}[b]{0.2\textwidth}
    			\includegraphics[width=\textwidth]{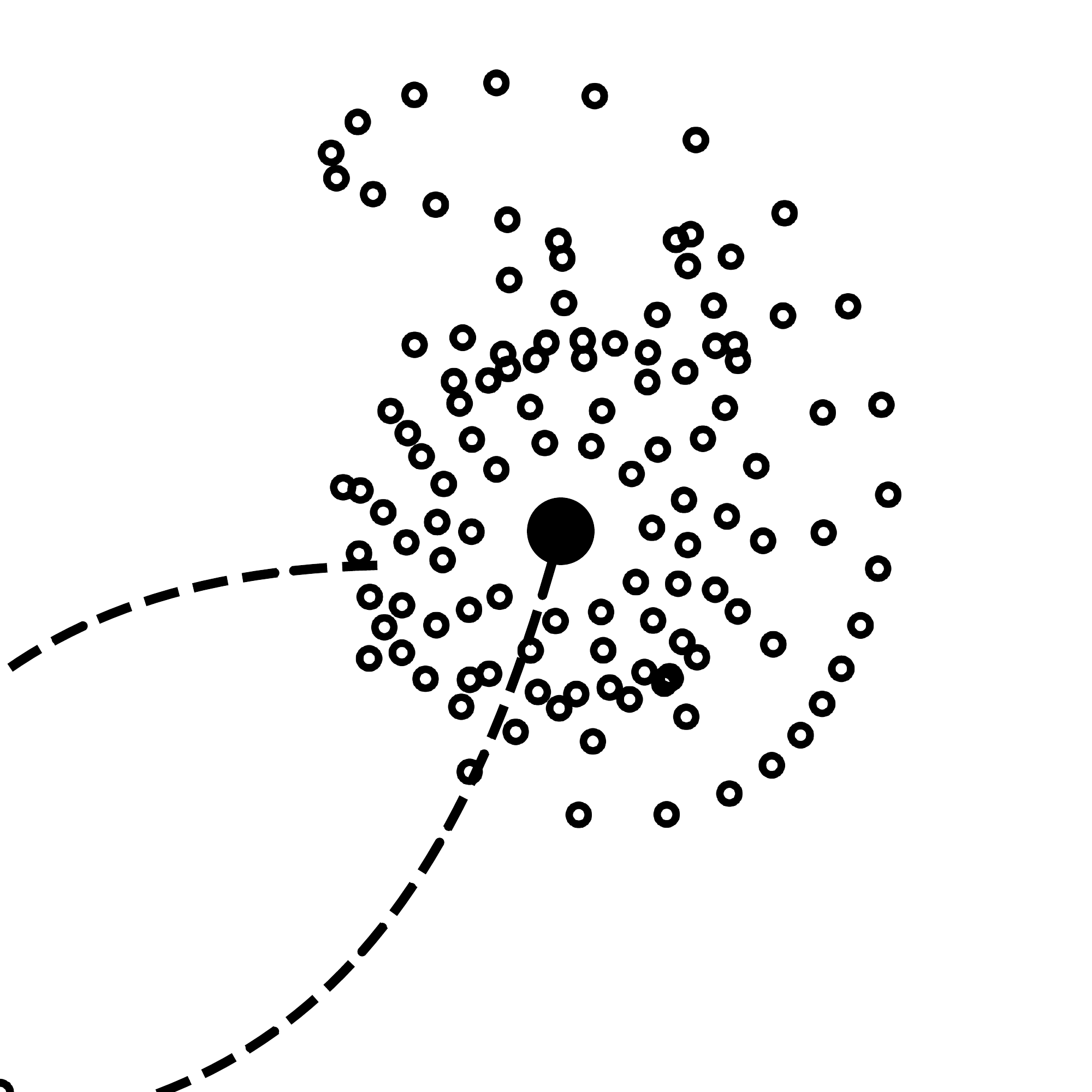}
    				\caption{$t = 6$}
  			\end{subfigure}
            \hfill
            \begin{subfigure}[b]{0.2\textwidth}
    			\includegraphics[width=\textwidth]{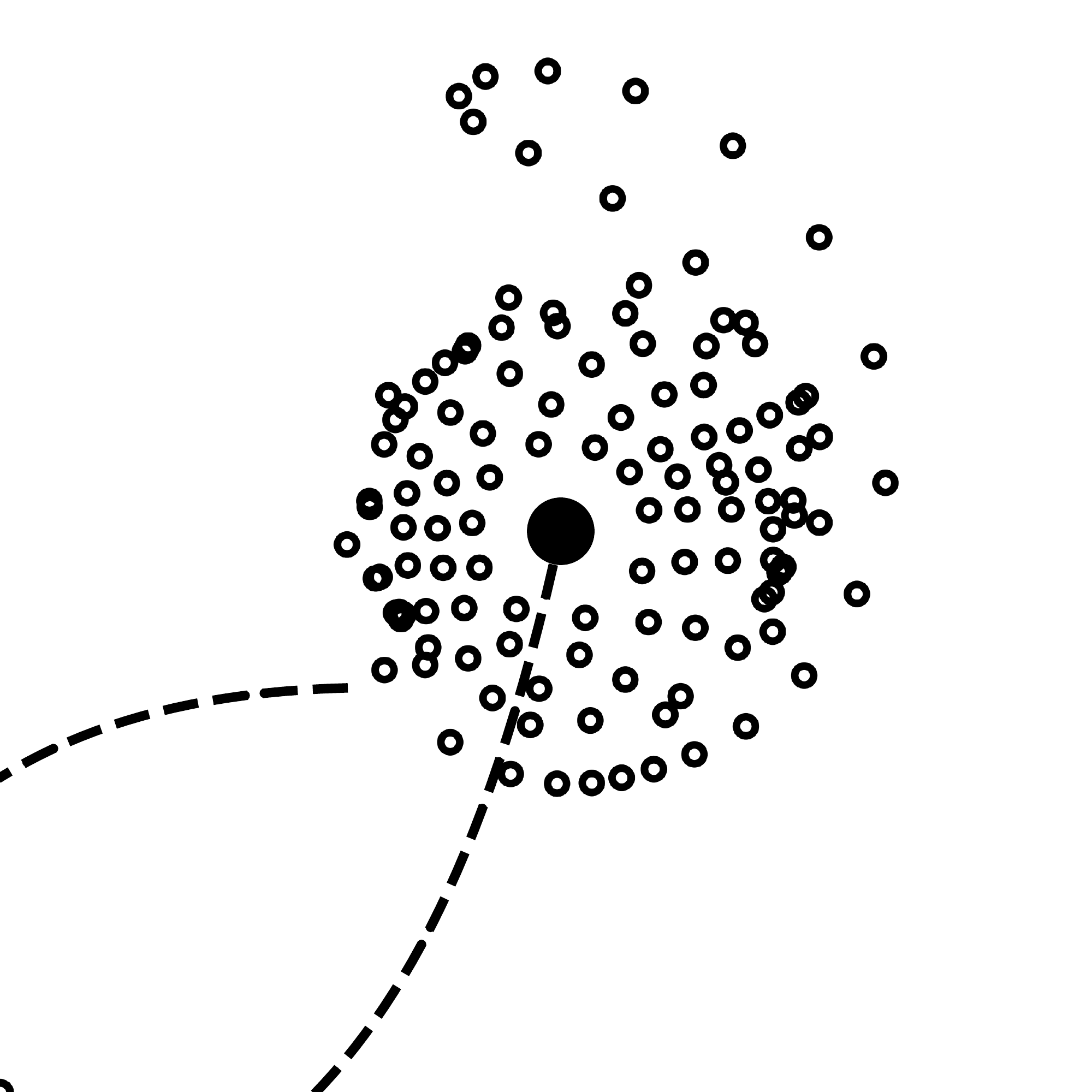}
    				\caption{$t = 7$}
  			\end{subfigure}
            \hfill
            \begin{subfigure}[b]{0.2\textwidth}
    			\includegraphics[width=\textwidth]{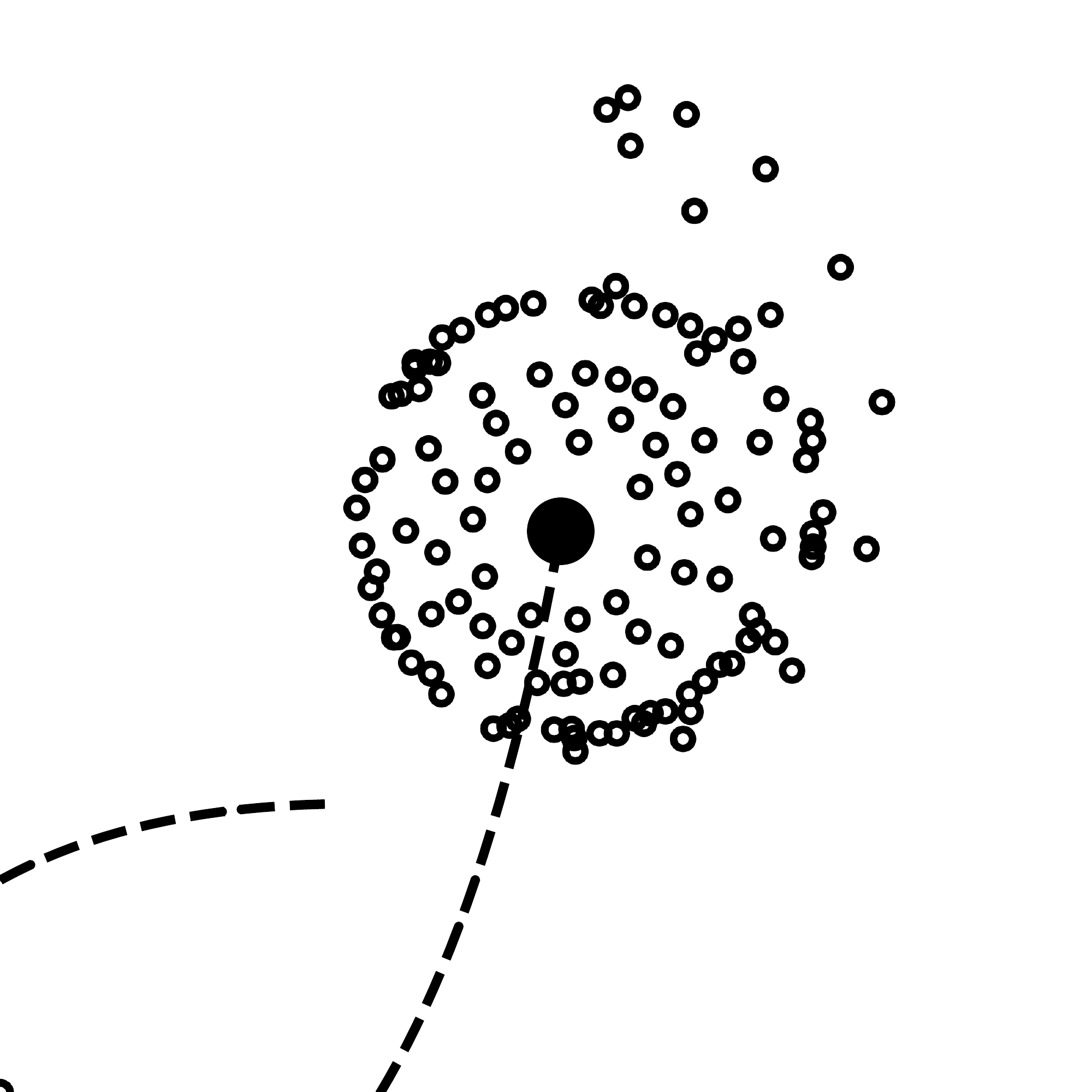}
    				\caption{$t = 8$}
  			\end{subfigure}
            \caption{Flat retrograde passage of a companion with equal mass, simulated with $120$ particles. Both the main galaxy and the companion have a mass of $10^{11}\: M_\odot$. All time values are given in multiples of $10^8\:$ years. In this example, $e \approx 1.21$ and $R_{min} \approx 30.46 \unit{kpc}$.}
            
            \label{fig:fig3}
		\end{figure}
Fig. \ref{fig:fig3} shows two galaxies with equal masses. One of the galaxies has a ring of test particles; we observe how those test particles react to the passage of the second galaxy. The orbit of the perturbing body is retrograde with respect to the revolution of the first galaxy's particles, in other words: the sense of rotation of the second galaxy's orbit is opposite to the sense of rotation of the test particles in the rings of the first galaxy.

As seen at $t = 3\cdot 10^8\:$years and onwards, only the outermost rings show obvious perturbations in their orbit. This results in the formation of a galactic tail at $t = 5\cdot 10^8\:$years, which dissolves soon after. The result is similar to that of Toomre \& Toomre  \cite{toomre1972galactic}. A small difference, namely the extent of the outermost regions of the disk at $t=5\cdot 10^9\:$years, is readily explained by differences in the initial values.
        
\subsection{Direct passage of a smaller companion}
        
\begin{figure*}[!htbp]
        \captionsetup[subfigure]{labelformat=empty}
  			\begin{subfigure}[b]{0.2\textwidth}
    			\includegraphics[width=\textwidth]{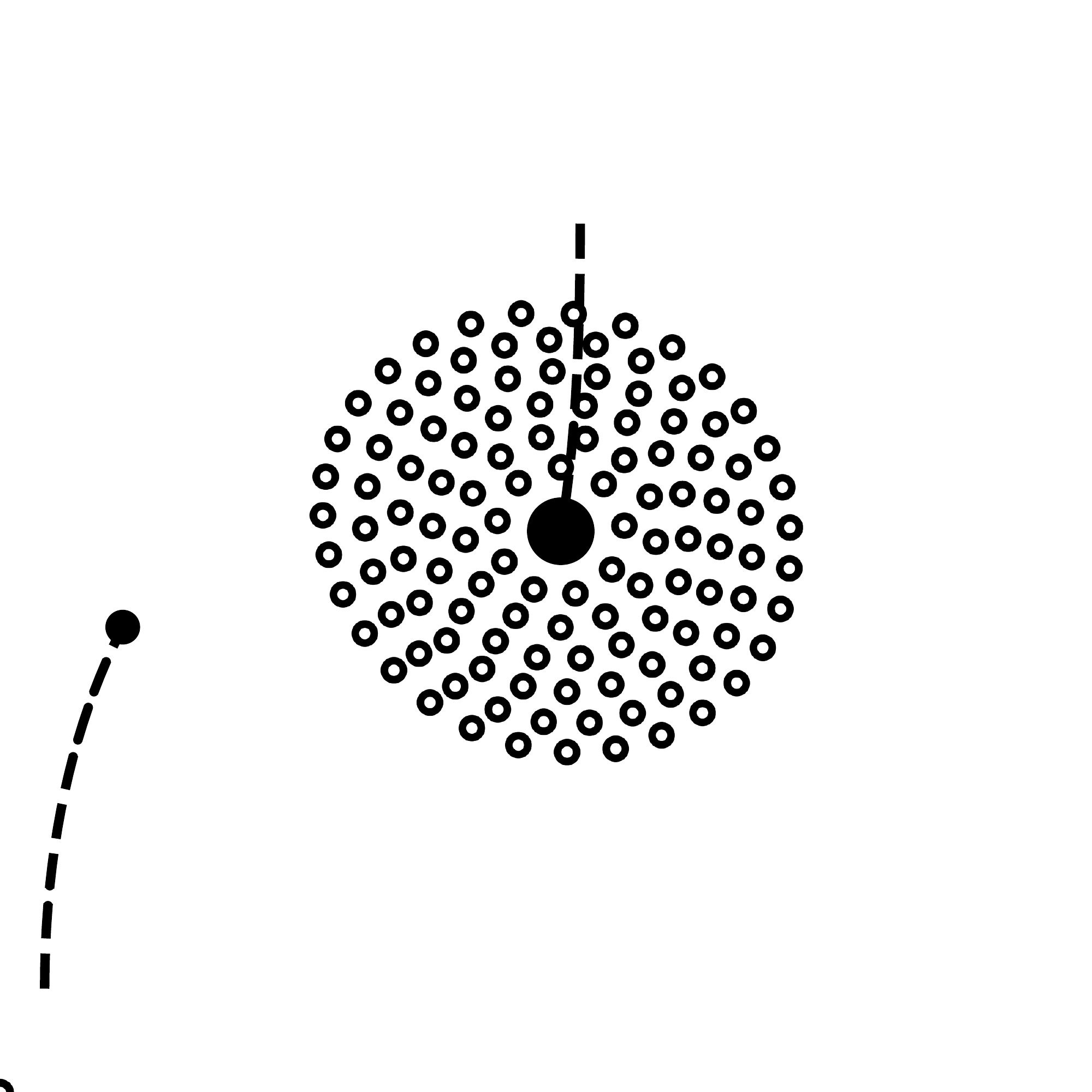}
    			\caption{$t = 4$}
  			\end{subfigure}
  			\hfill
  			\begin{subfigure}[b]{0.2\textwidth}
    			\includegraphics[width=\textwidth]{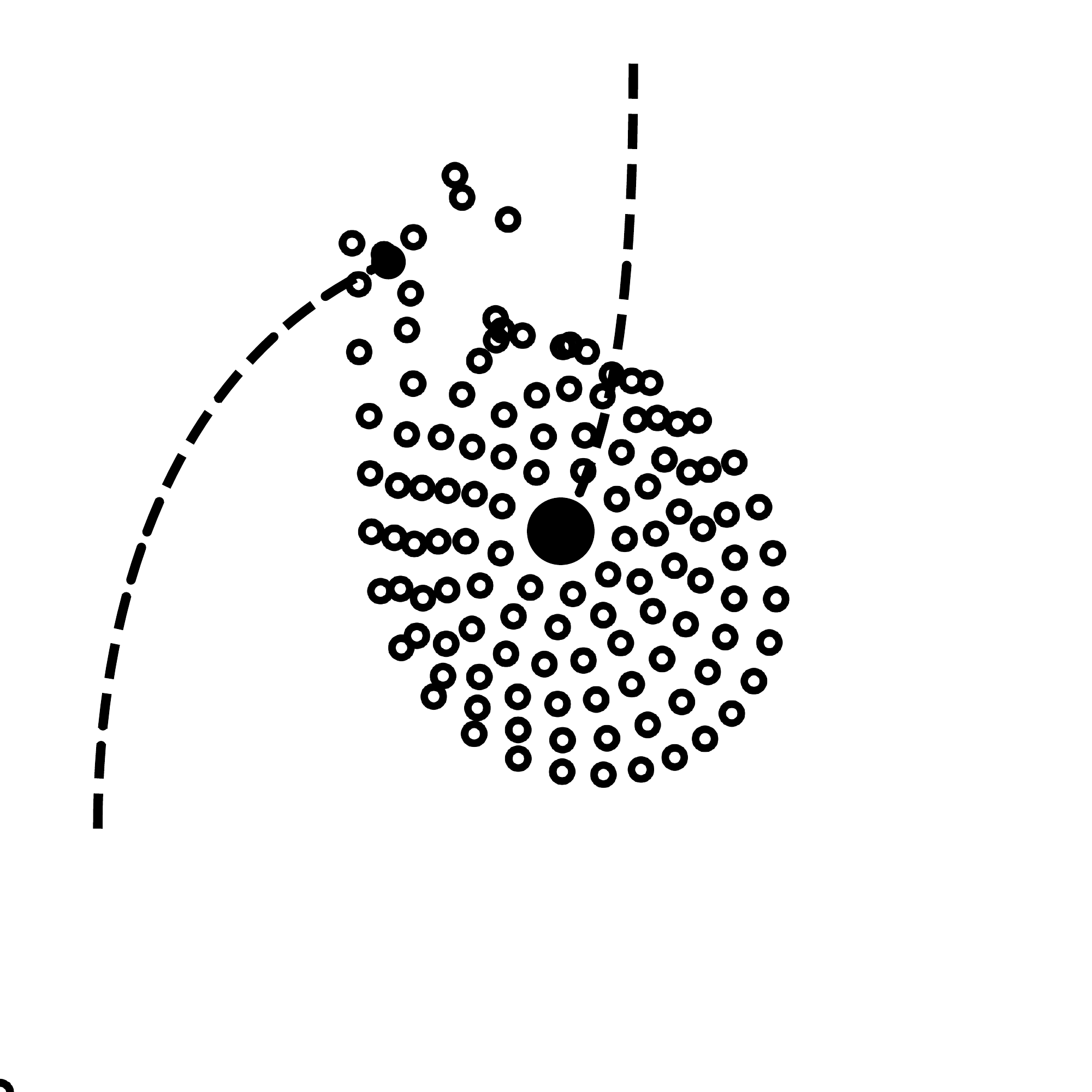}
    				\caption{$t = 6$}
  			\end{subfigure}
            \hfill
            \begin{subfigure}[b]{0.2\textwidth}
    			\includegraphics[width=\textwidth]{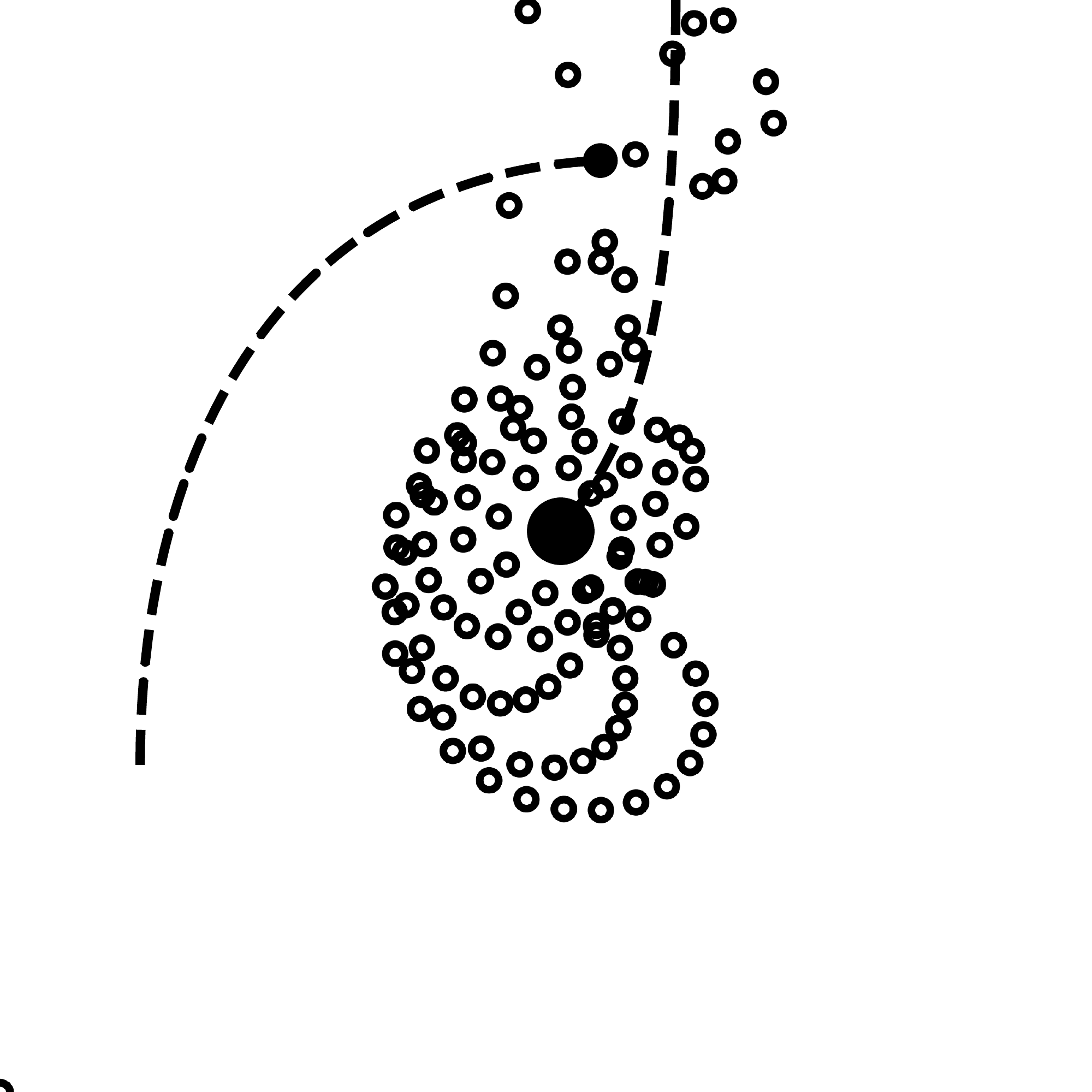}
    				\caption{$t = 7$}
  			\end{subfigure}
            \hfill
            \begin{subfigure}[b]{0.2\textwidth}
    			\includegraphics[width=\textwidth]{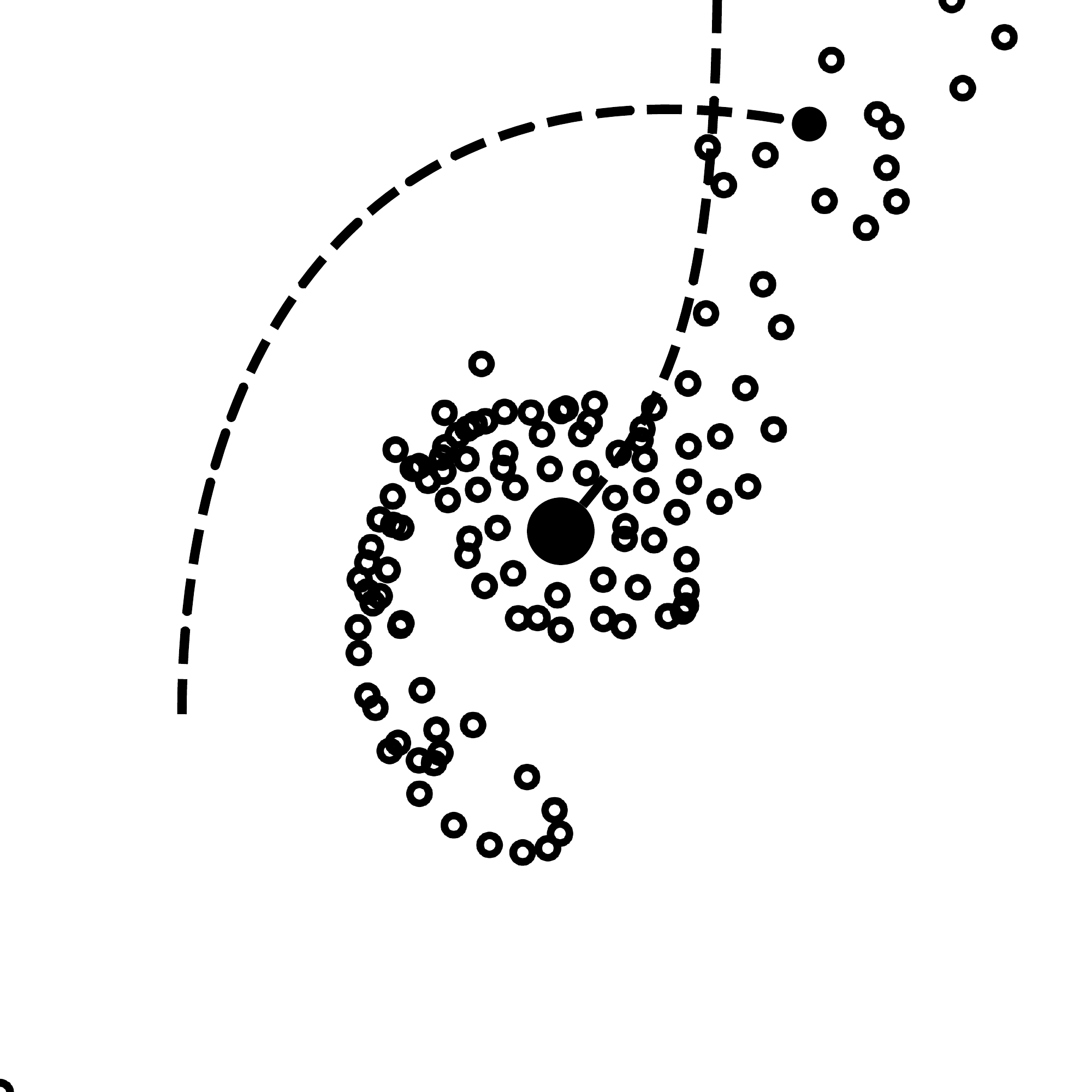}
    				\caption{$t = 8$}
  			\end{subfigure}
  			
            \begin{subfigure}[b]{0.2\textwidth}
    			\includegraphics[width=\textwidth]{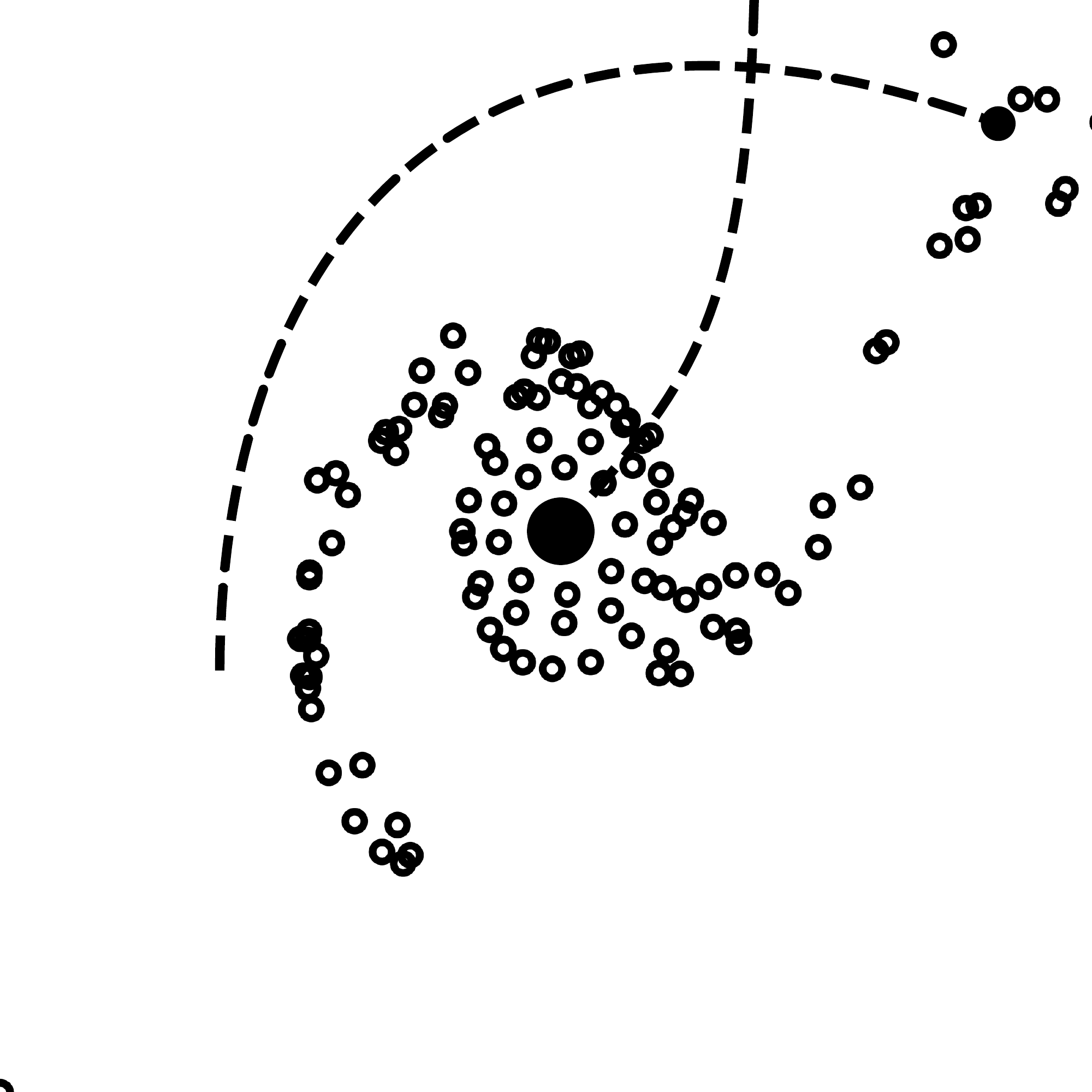}
    				\caption{$t = 9$}
  			\end{subfigure}
            \hfill
            \begin{subfigure}[b]{0.2\textwidth}
    			\includegraphics[width=\textwidth]{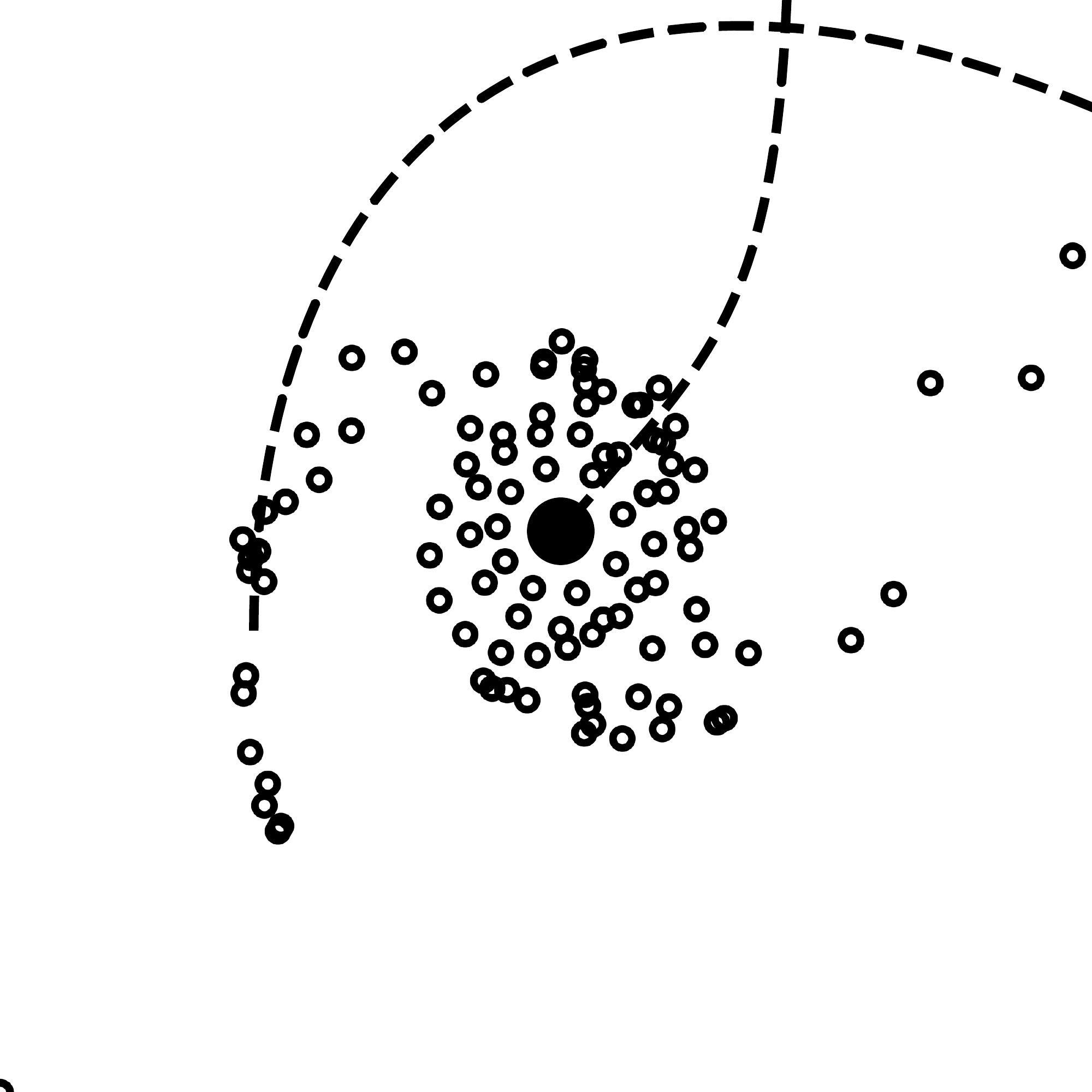}
    				\caption{$t = 10$}
  			\end{subfigure}
            \hfill
            \begin{subfigure}[b]{0.2\textwidth}
    			\includegraphics[width=\textwidth]{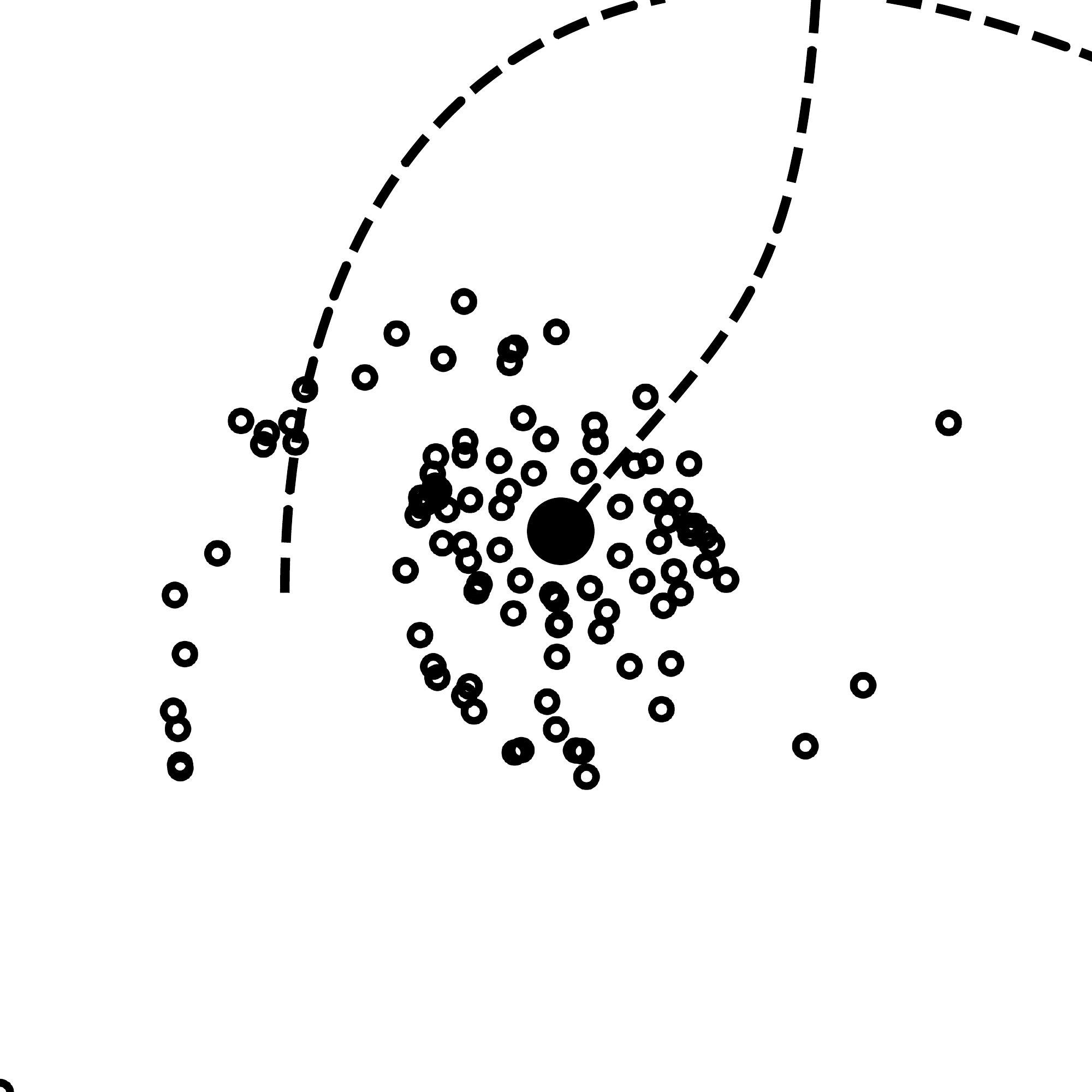}
    				\caption{$t = 11$}
  			\end{subfigure}
            \hfill
            \begin{subfigure}[b]{0.2\textwidth}
    			\includegraphics[width=\textwidth]{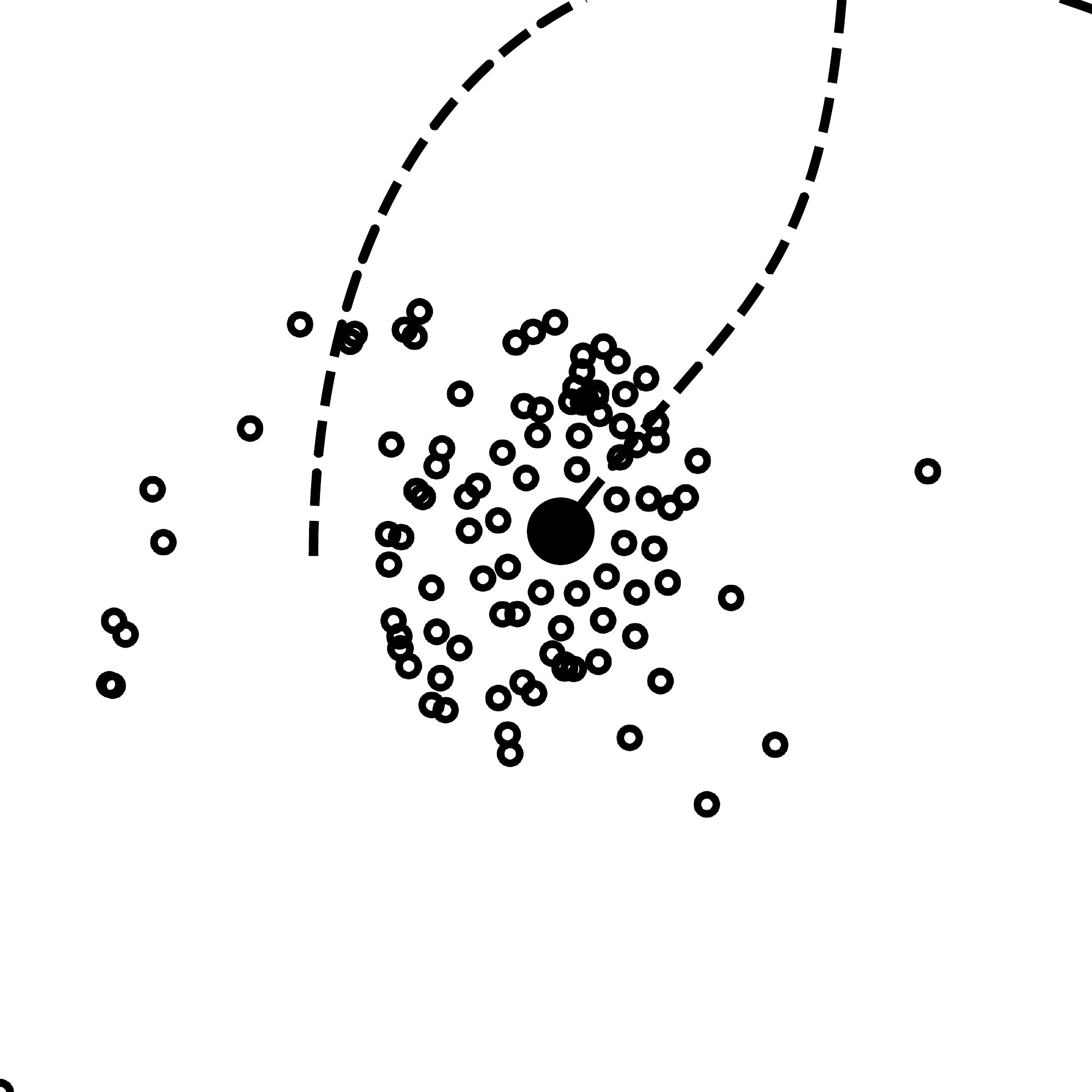}
    			\caption{$t = 12$}
                \label{fig:fig4h}
  			\end{subfigure}

            \caption{Flat direct passage of a smaller companion, simulated with $120$ particles: main galaxy mass $10^{11}\: M_\odot$, companion galaxy mass $2.5 \cdot 10^{10}\: M_\odot$. All time values are in multiples of $10^8\:$ years.
            The orbital parameters are $e \approx 1.04$ and $R_{min} \approx 27.02\unit{kpc}$.} 
                        \label{fig:fig4}
		\end{figure*}

        In this example, we are again looking at a galaxy with test-particle ring. This time, the passing companion has a mass only a quarter that of the first galaxy. Furthermore, this is a \textit{direct} passage: the sense of rotation of the passing companion's orbit relative to the first galaxy is the same as the sense of rotation of the test particles in the first galaxy's rings. The resulting encounter is shown in Fig. \ref{fig:fig4}. A prominent feature of this encounter is the formation of a bridge between the main galaxy and the companion. The bridge does not last very long, however. By the time the simulation has reached about $t=12\cdot 10^8\:$years, this bridge has dissipated completely. Notably, this means the bridge is much shorter-lived than the tidal counter-arm that has also been produced in the encounter.                  

    \subsection{Three-dimensional model of NGC 5426}

\begin{figure}[htbp]
\begin{center}
\includegraphics[width=\linewidth]{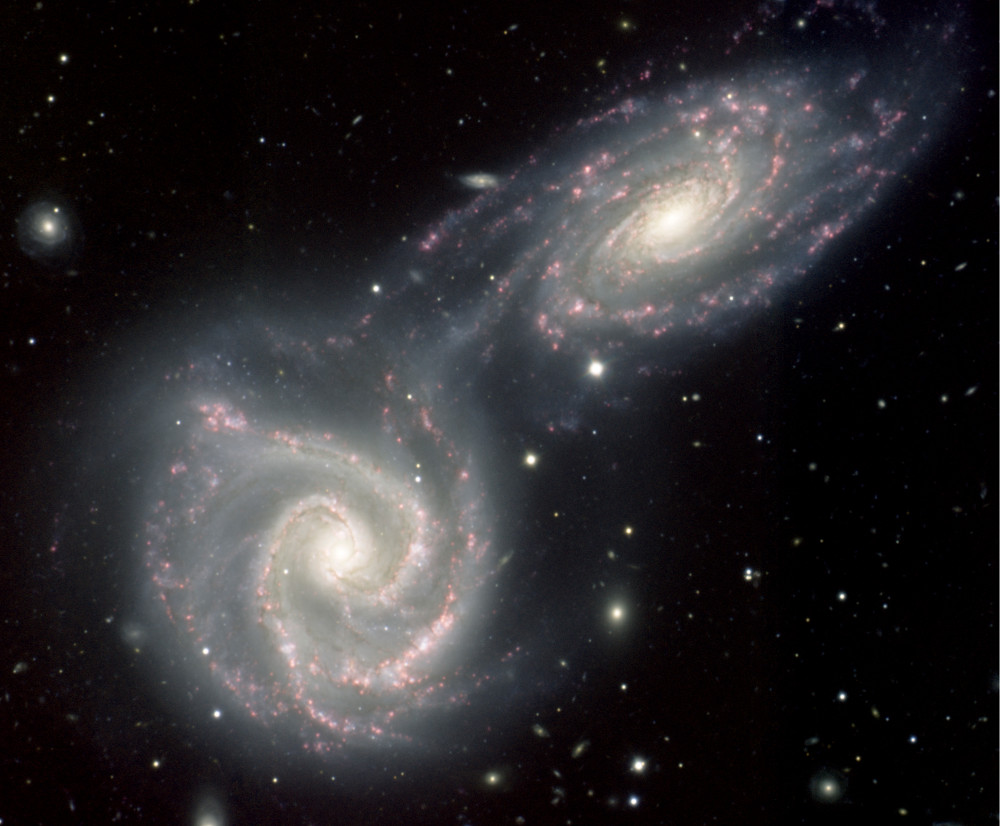}
\caption{The galaxy pair NGC 5426 and 5427. Image credit: Gemini Observatory / NRC/ AURA / Christian Marois et al. [Gemini Observatory Legacy Image]}
\label{NGC5426n7}
\end{center}
\end{figure}

The galaxy pair consisting of NGC 5426 and NGC 5427, which is shown in Fig. \ref{NGC5426n7}, is a remarkable example of spiral arms and a bridge forming due to tidal interactions. As part of our project, we attempted to recreate these structures. 
        
In the encounter we simulated, the revolution of the particles were retrograde to the orbit of the interacting galaxies with each other. The results of our simulation can be seen in Fig. \ref{fig:ngc5426}. From the chosen perspective, NGC 5426 has initially started on the left side of the view. By the time $t=7\cdot 10^8\:$years, the first close encounter has already occurred, and NGC 5426 can be seen to the right of its companion.

In the snapshot at  $t=7\cdot 10^8\:$years, one can see the bridge connecting the two galaxies. 
        
In the snapshots at $t=7\cdot 10^8\:$years and $t=12\cdot 10^8\:$years, both galaxies are moving away from each other, and are approaching their apocenters by $t=22\cdot 10^8\:$ years.
     
At $t=22\cdot 10^8\:$years the simulated galaxies share visual similarities with their observed counterparts: global spiral arms similar to those in the observed galaxies are present.
        
        \begin{figure*}[!tbp]
        \captionsetup[subfigure]{labelformat=empty}
        	
            \begin{subfigure}[b]{0.45\textwidth}
    			\includegraphics[width=\textwidth, trim={1cm 6cm 1cm 6cm}, clip]{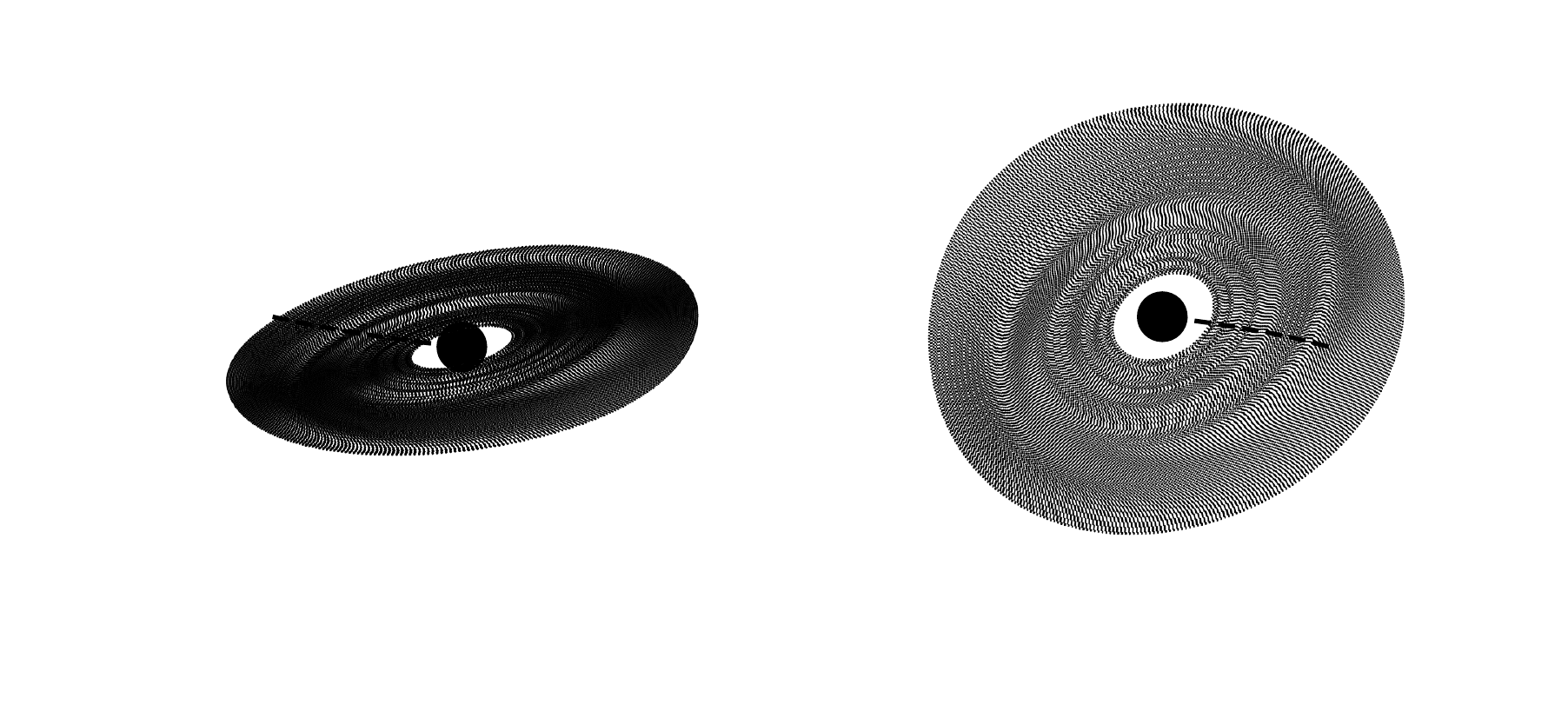}
    			\caption{$t = 2$}
  			\end{subfigure}
            \hfill
            \begin{subfigure}[b]{0.45\textwidth}
    			\includegraphics[width=\textwidth, trim={1cm 6cm 1cm 6cm}, clip]{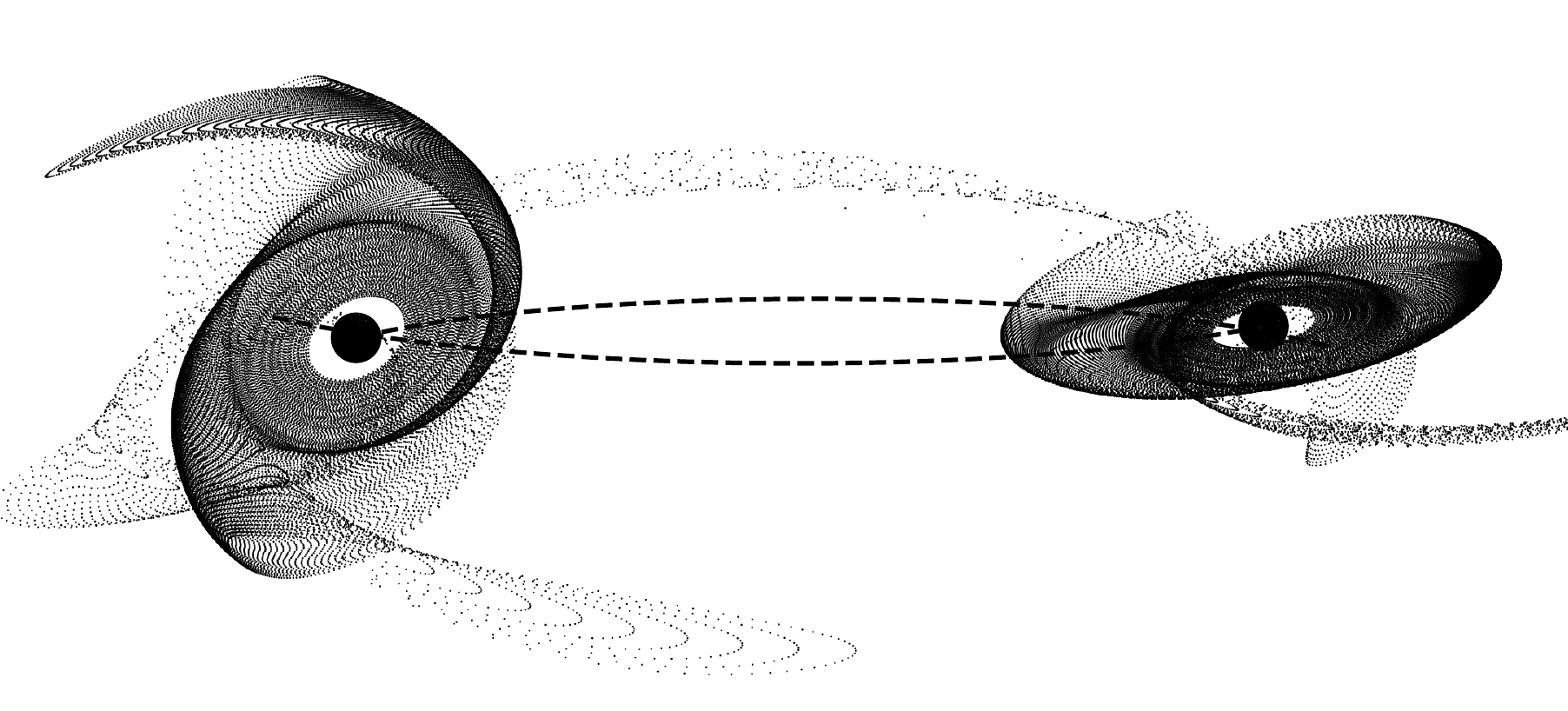}
    			\caption{$t = 7$}
  			\end{subfigure}
            \centering
            \begin{subfigure}[b]{0.9\textwidth}
    			\includegraphics[width=\textwidth, trim={0cm 5.5cm 0cm 5.5cm}, clip]{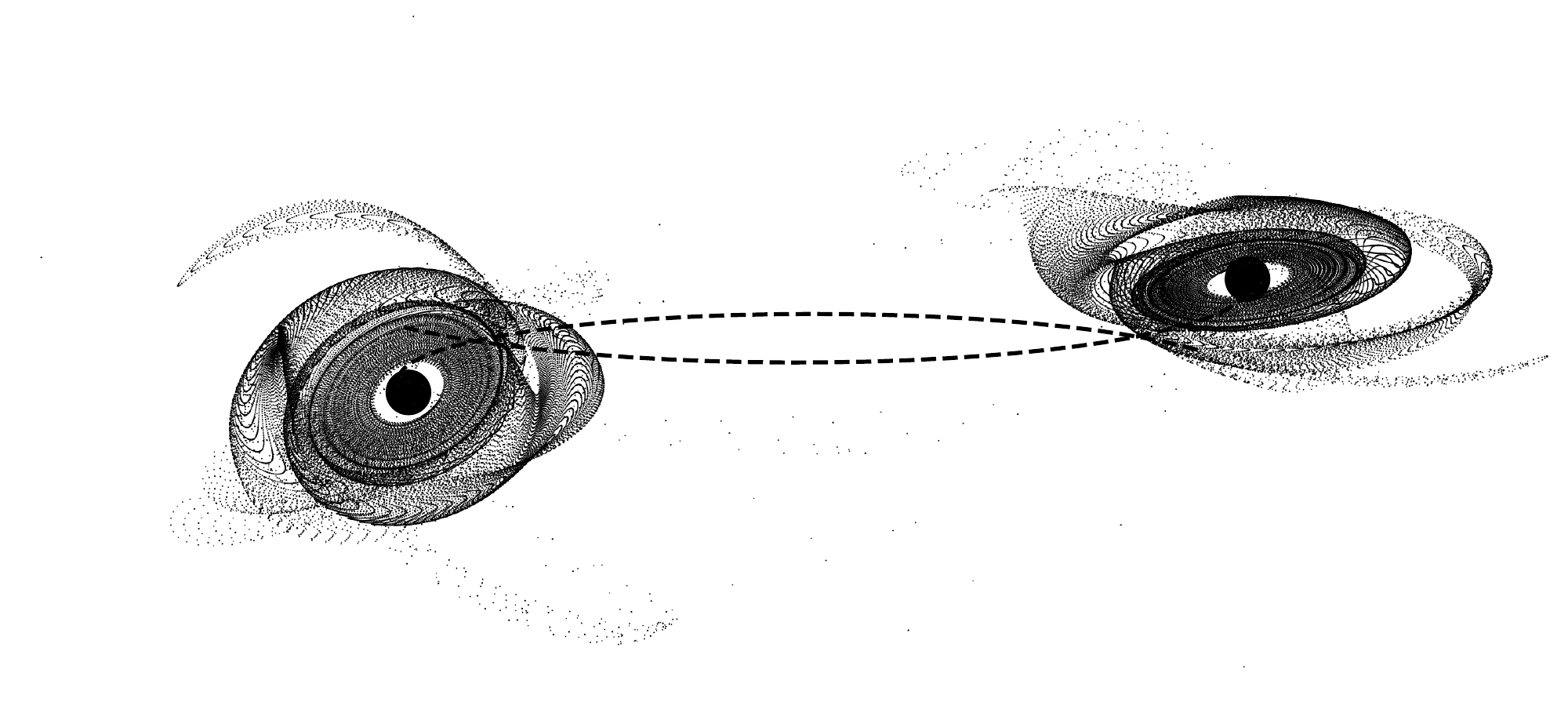}
    			\caption{$t = 12$}
  			\end{subfigure}
			
            \begin{subfigure}[b]{0.9\textwidth}
\includegraphics[width=\textwidth, trim={0cm 5.5cm 0cm 4cm}, clip]{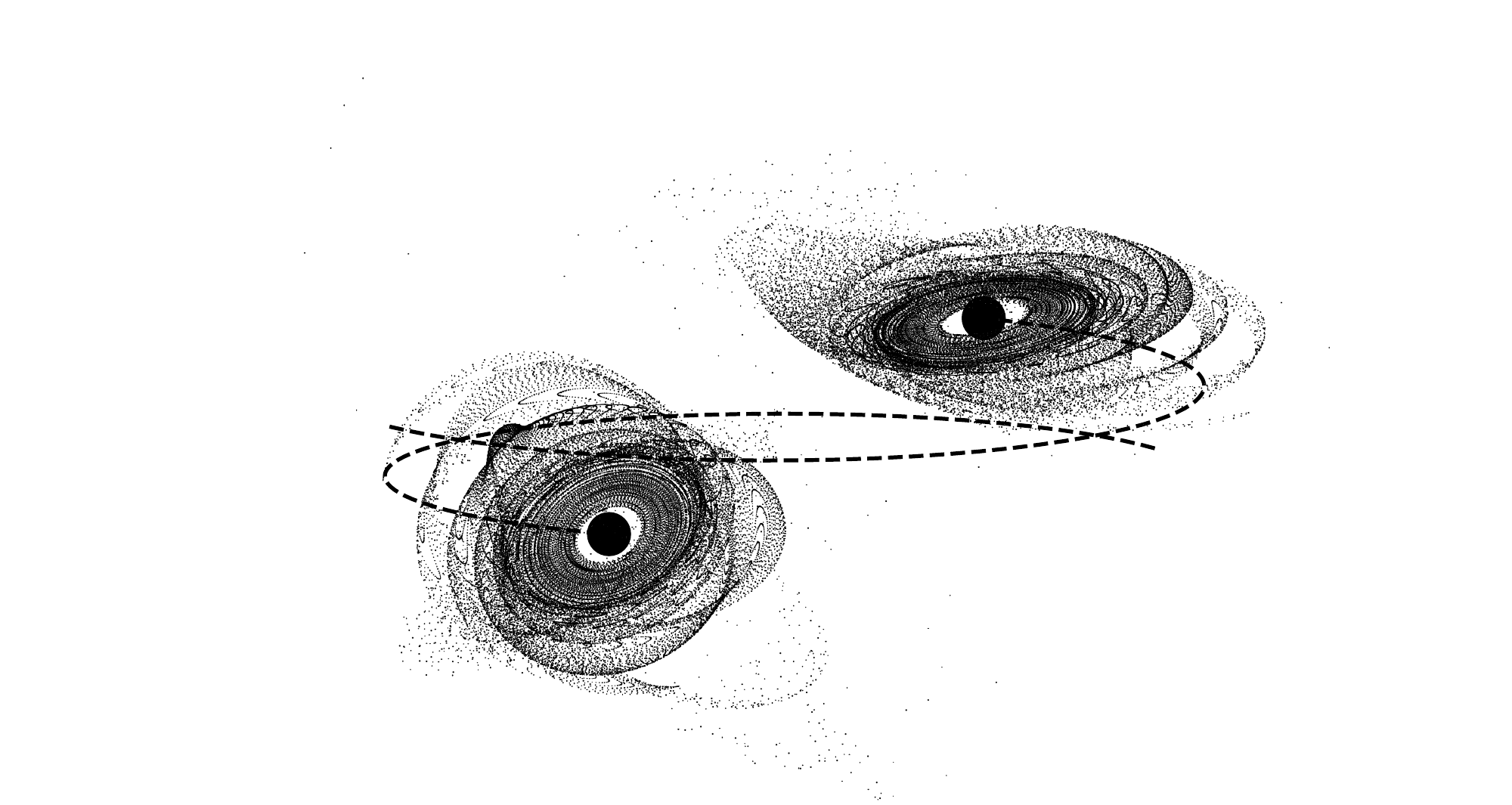}
    			\caption{$t = 22$}
  			\end{subfigure}
            
            \caption{Simulated interaction between galaxies analogous to NGC 5426 and NGC 5427, with a total of $75~000$ particles ($37~500$ in each galaxy's initial disk). The encounter was a retrograde passage where both galaxies had equal mass. The projection plane formed and angle of $83^{\circ}$ with the two central particles' orbital plane. The two particle disks have a relative inclination of $30^{\circ}$ with respect to each other. In this simulation, the eccentricity of the orbit is $e \approx 0.67$ and the distance of closest approach is $R_{min} \approx 28.16 \unit{kpc}$.}
            \label{fig:ngc5426}
            
            \label{fig:ngc5426}
		\end{figure*}

        \subsection{Antennae galaxies NGC 4038/9}
    \begin{figure}[htbp]
\begin{center}
\includegraphics[width=\linewidth]{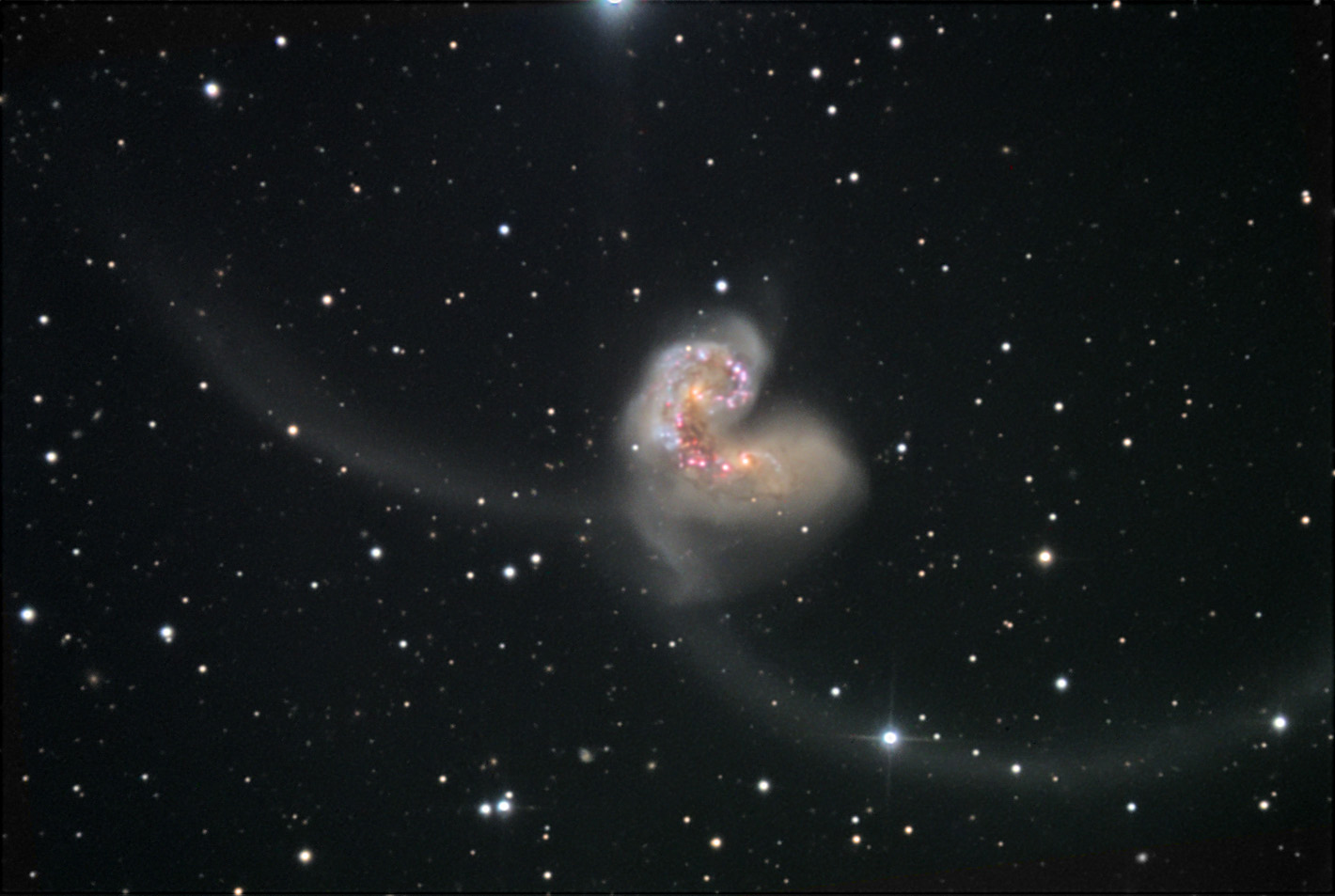}
\caption{Visible-light observation of the antennae galaxies NGC 4038 and NGC 4039. Image credit: Bob and Bill Twardy/Adam Block/NOAO/AURA/NSF}
\label{AntennaeKP}
\end{center}
\end{figure}

         \begin{figure*}[!tbp]
         \captionsetup[subfigure]{labelformat=empty}


			\centering
            \begin{subfigure}[b]{0.9\textwidth}
            \includegraphics[width=\textwidth, trim={0 4cm 0 6cm}, clip]{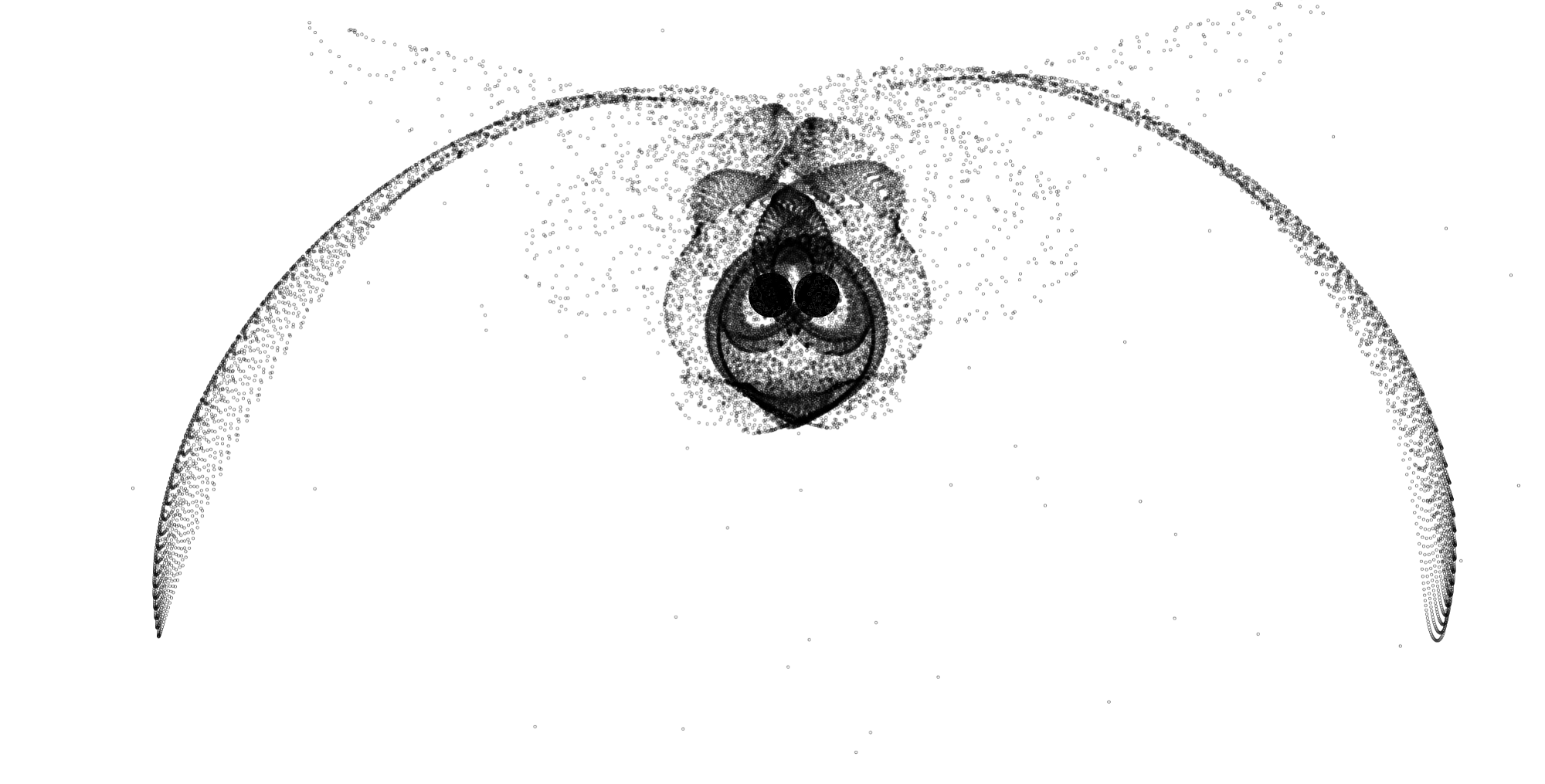}
    		
  			\end{subfigure}

            \begin{subfigure}[b]{0.9\textwidth}
            \includegraphics[width=\textwidth, trim={0 3cm 0 3cm}, clip]{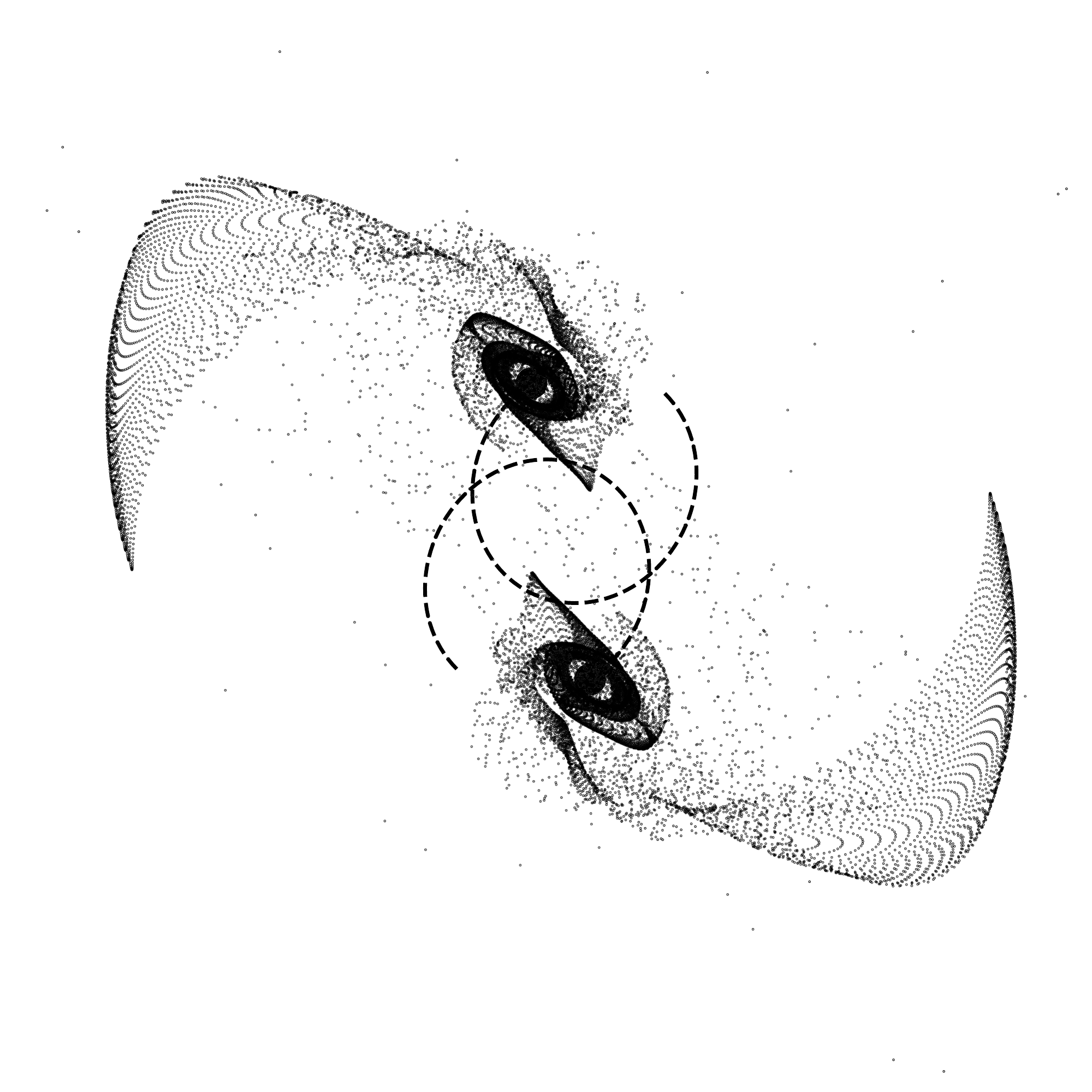}
    			\caption{$t = 9$} 
  			\end{subfigure}

            \caption{Simulated interaction between NGC 4038 and NGC 4039, commonly known as "The Antennae", with a total of $25~000$ particles ($12~500$ within each galaxy's initial disk). The encounter was a direct passage where both galaxies had equal mass. In the first subfigure, the projection plane is perpendicular to the orbit plane. However, in the second depiction of the galaxies, the angle formed by the orbit plane and the projection plane is $60^{\circ}$. In this passage, $e \approx 0.44$ and $R_{min} \approx 18.57 \unit{kpc}$.}

            \label{fig:antennae}
		\end{figure*}
        
        The last simulation presented here is an attempt to replicate the encounter that gave rise to the antennae-like shape of NGC 4038/9. A view of the antennae galaxies, taken with the $20''$ telescope at Kitt Peak Observatory, can be seen in Fig. \ref{AntennaeKP}. For comparison, the results of the final version of our simulation are shown in Fig.  \ref{fig:antennae}.
  
        Using an eccentricity value similar to Toomre \& Toomre's ($e = 0.44$ instead of $e=0.5$), the resulting shape closely resembles the structure of the observed pair. Fig. \ref{fig:antennae}'s second depiction of $t=9$, where the projection plane is no longer perpendicular to the orbit plane, allows for a better understanding of the relative positions of the two galaxies. In that same depiction, it is made clear that there is no actual crossing of the tails; their symmetrical shape resembles Toomre \& Toomre's Fig. 23, albeit with some noticeable differences: in our example, the tails are considerably wider and many particles have been extracted from the disks with an apparently arbitrary velocity. 
        
        A likely explanation for those differences is that for modelling the Antennae, Toomre \& Toomre introduce a modified, ``softened'' gravitational potential $-GM/(r^2 + a^2)^{1/2}$, where they set $a = 0.2R_{min}$ in order to mimic the affect of an extended mass (as opposed to a point mass). One result of this, as they noted, was indeed a reduced width of the tails.

  \label{EndProject}     
                  
    \section{Lessons learned, and recommendations} \label{education}

Personalized accounts of the experiences of M. B.--C. and M. T. can be found in appendices \ref{Brea} and \ref{Thiel}, respectively. This section is our attempt to extract more general lessons we learned from the project, which we believe to be of relevance to those intending to introduce a project of this kind in a high-school setting. 
     
In terms of prerequisites, we found that in our case, the most important ingredient on our part was solid interest in the project. Students replicating our project as a whole --- that is, writing all of their own code --- will undoubtedly face obstacles while working and spend numerous hours troubleshooting their code. They will feel frustrated at more than one point. That is where interest and commitment become particularly important for helping students to stay the course. We have found that with a solid level of motivation and commitment, plus the required concepts from high school physics and maths, reaching the proposed goals is feasible; in addition, some basic linear algebra was a big plus for the optimization strategy discussed in section \ref{Optimization}. 

In the form presented here, this project requires a considerable amount of time; as such, it could be suitable as a programming project linking the subjects of physics and computation (such as the newly introduced subject IMP linking computer science, mathematics, and physics, in the German state of Baden-W{\"u}rttemberg).\footnote{
Further information, in German, can be found on the official site at \href{http://www.bildungsplaene-bw.de/,Lde/LS/BP2016BW/ALLG/SEK1/IMP}{http://www.bildungsplaene-bw.de/,Lde/LS/BP2016BW/ALLG/SEK1/IMP} (web pages last accessed on 18 August, 2018).
} Given that the project provides opportunities for introducing most of the key programming concepts, such as list or array structures, simple calculations and evaluations of expressions, conditions, and loop structures, it could be paired with a general introduction to these concepts in the framework of an introductory programming class. 

The students could work their way up to the full numerical integration using simpler examples such as the harmonic oscillator, similar to our on exploration of numerical stability as described in section \ref{StabilitySection}. Starting with two-dimensional situations provides a further simplification step. 

Conceptually, we found the Euler method easiest to understand, which already provides a useful first numerical integration scheme. For advanced students, who have a notion of the concept of a derivative as an infinitesimal approximation of a function's graph, the connection between that approximation and Euler integration can be made explicit.

Once students have programmed a working two-dimensional orbit simulator on their own, they are prepared to use a paper such as the one of Toomre \& Toomre \cite{toomre1972galactic}, and should be able to tackle replications of those authors' ''Four Elementary Examples'', all of which are two-dimensional.

A reduced version of our project might well end here. By this point, students should have achieved a basic understanding of numerical integration and orbital mechanics in a two-dimensional space. An optional, but useful further step would be the introduction of the velocity Verlet algorithm, together with an elementary discussion of numerical stability similar to the one in our section  \ref{StabilitySection}.
  
The next major step is the transition from a two-dimensional to a three-dimensional simulation. Judging by our own experience, the most difficult aspect of this step is the description of the geometry of this situation in terms of the standard orbital elements. This requires spatial thinking, and a mental map from the three-dimensional situation to the triangles and their angles used to parametrize the orbit. Personally, we found Rodrigues' rotation formula  \cite{GrayRodrigues} to be a helpful short-cut. 
  
Once this transition is achieved, students can tackle the task of using their simulations to recreate real astronomical images of interacting galaxies, similar to what we reported on in our section \ref{simulations}. Examples for interacting galaxies suitable for this part of the project, for which images are readily available online, can be found in table \ref{InteractingGalaxies}.
\begin{table}[htp]
\begin{center}
\bgroup
\renewcommand{\arraystretch}{1.2}
\begin{tabular}{|c|}
\hline
Galaxy designation\\\hline\hline
\href{https://www.spacetelescope.org/images/heic0812c/}{NGC 4038/4039 (''Antennae'')}\\\hline
\href{https://www.spacetelescope.org/images/heic0206h/}{NGC 4676 (''The mice'')}\\\hline
\href{http://hubblesite.org/image/1627/news_release/2004-45}{NGC 2207 and IC 2163}\\\hline
\href{https://www.spacetelescope.org/images/heic0810bv/}{AM 0500-620}\\\hline
\href{https://apod.nasa.gov/apod/ap130825.html}{Arp 271}\\\hline
\href{https://www.spacetelescope.org/images/heic0810am/}{NGC 6786}\\\hline
\href{https://www.spacetelescope.org/images/heic0810af/}{UGC 9618}\\\hline
\href{https://www.spacetelescope.org/images/heic0810al/}{UGC 8335}\\\hline
\href{https://www.spacetelescope.org/images/heic0810ag/}{Arp 256}\\\hline
\href{https://www.spacetelescope.org/images/heic0810aj/}{ESO 593-8}\\\hline
\href{https://www.spacetelescope.org/images/heic0810ao/}{ESO 77-14}\\\hline
\end{tabular}
\egroup
\end{center}
\caption{List of notable interacting galaxies}
\label{InteractingGalaxies}
\end{table}%

Our description so far refers to a version where the students work comparatively independently. For us (M. B.-C. and M. T.), one of the most positive aspects of this project was that we were often left with little to no guidance, forcing us to do our own research and develop our problem-solving skills. We suggest that a similar approach could be beneficial when recreating this project with suitably advanced and motivated students.

In addition, versions of this project with reduced levels of difficulty can be envisioned (albeit at the expense of also losing the corresponding benefits of independent work). For instance, just as scientists frequently learn new software skills by studying and by creating variations of existing script and programs, students can be given existing scripts, or parts of such scripts, to experiment with, or to add specific elements that have deliberately been left out. The scripts written in the course of this project, available at GitHub\footnote{\href{https://github.com/mbrea-c/simulating-tidal-interactions.git}{https://github.com/mbrea-c/simulating-tidal-interactions.git}}, can be used for this purpose.

\section*{Acknowledgements}
    
    The work described in this paper was only made possible through this internship, and thus M. B.--C. and M. T.  sincerely thank Haus der Astronomie as well as the Max-Planck-Institute for Astronomy for offering this opportunity to high school students. Furthermore, they would like to express our gratitude towards both facilities for introducing them to the world of scientific research and providing insight into life as an astronomer. Lastly and most importantly, M. B.--C. and M. T. thank M. P. for supervising the internship and giving them the chance to attend the WE Heraeus Summer School and present their results. We thank Markus Nielbock for helpful comments on an earlier version of this manuscript.
    
    \clearpage
    
    \bibliography{galaxien-simulation}{}
    \bibliographystyle{plain}
    
    \clearpage
    \begin{appendix}
    \begin{strip}
   
    \section{Individual accounts by the two interns}
    
Since the two student interns, M.~B.-C. and M.~T., come from considerably different backgrounds, their experiences throughout the internship were significantly different, and each has written a separate personal account documented in this appendix.

\subsection{Manuel Brea-Carreras} \label{Brea}
It should be noted that I started working in this project having programmed a few times in the past; not enough to make me an experienced programmer by any means, but enough so that I did not have to go through the hoops of learning to code for the first time. Therefore, my personal advice in this case is not directed to helping students get through those first steps in programming but instead to those who, having coded before, still lack experience working on larger projects with a definite goal. 

This is not to say that previous programming experience is required to get anything out of this project; in fact, M.~T. did start learning Python from scratch in the first three weeks of the internship, and I found myself having to catch up to his progress when I joined for the last three weeks before we teamed up and developed this project to the extent that is presented in this paper.

More experienced programmers will understand the importance of thinking through a project's structure before writing the first line of code, specially if the project is going to require more than a few days' worth of work. As an inexperienced student and having joined the internship only for the last three weeks, I was too eager to get a working prototype of the simulation of my own, and I overlooked many aspects of the design that would cause issues later on. In retrospect, I would advise other students to spend a reasonable amount of time thinking of and discussing a modular and easy-to-work-on overall structure before starting to type. As a next step, it might be helpful to write the skeleton of the project before starting to fill in the functions.

One specific bad design choice that I would like to point out is our attempted Object-Oriented structure. Even though Object-Oriented Programming undeniably has many potential use cases, a bad implementation caused by a lack of experience can ---and will--- hinder future development. In our case, the OOP implementation I proposed early on turned to be an obstacle in the way of optimization later on in the project and forced us to waste time removing it and refactoring much of the code. My advice to those who only know the basics of OOP is to refrain from using the paradigm, unless training on its application is part of your goal.

Finally, another issue that I think we could have handled better is time management; our goal was not simply to write the code for the simulations, but also a report on the results and, later on, prepare a presentation for a conference. However, we ended up spending a disproportionate amount of time in the former, and we consequently had to compile most of the results we presented in a small window of time ahead of our presentation at the Heraeus Summer School. My advice, again, is for students to think before they type; you should consider whether that new idea is going to provide any utility towards your goal with this project, or it is just unnecessary eye-candy for the simulation output.

\subsection{Michael Thiel} \label{Thiel}

As mentioned above, I had only little experience in programming before starting the internship. While I was somewhat familiar with basic concepts found in programming languages (like variable declarations and loop structures), I had never touched on more advanced aspects of it let alone written an actual program. I therefore had to learn Python from scratch before I could begin working on the actual project. Through research, some programming exercises and an online course,{\footnotemark} I was able to pick up most of what was needed start working by the beginning of the second week. 
Although I certainly enjoyed working on the project throughout its entirety, there were times of frustration I had to face. Especially in the early stages, when I was still familiarizing myself with Python, I was often unhappy with my rate of progress. Even though I learned a great deal about programming at the time, there was not much progress being made in terms of the actual project.

A strategy I followed in an attempt to limit this frustration was to set myself small goals while learning Python. By not focusing on the completion of the whole project during the early stages, and instead concentrating on one aim at a time, I found it a lot easier to stay dedicated. Working on incrementally more complex and challenging goals also made progress more visible and the project scope seem less overwhelming. I advise students who find themselves in in a similar position and become discouraged, to divide their project into chunks and measure progress in a comparable manner. I had a lot of success in avoiding unnecessary frustration with this approach and would go on and stick to it for the rest of the project. 

To give some more detail about the individual steps I took with this approach: After familiarizing myself with Python, I started out by simulating the motion of a simple one-dimensional harmonic oscillator as my first goal. Although it may seem unrelated to the project's topic, I learned most of the basics about simulations working on this scenario. My next aim, with the intent of getting closer to the actual project, was to simulate the orbit of a planet around a star. Most of what was needed in terms of programming structure I had already dealt with previously, when simulating the oscillator, so I spent the majority of my time implementing the proper orbital mechanics. Once this was finished, scaling the system up to a whole galaxy was less of an obstacle than I had previously imagined. The physics remained unchanged, as I had decided to do calculations numerically from the very beginning. Thus, the transition was mostly a matter of adding more bodies and some extra code (to ensure their correct placement for given parameters) to make a whole galaxy.

Two weeks after starting the internship, I had a first working prototype script that could compute 2D encounters between a galaxy and a bare companion. The last week was spent refining the existing code, switching to the more stable velocity Verlet algorithm for numerical integration and replicating some of Toomre and Toomre's elementary examples. After I partnered up with M.~B.-C in the following week, the project progressed rapidly with our efforts combined, reaching a level that I could have never achieved by working on my own. For this reason, I can only recommend other students to work together as well, especially when the available time is strictly limited.

Altogether, by keeping some of this advice and a few of the mistakes we made in mind, students should be able to work through this or a similar project successfully.

\end{strip}

\footnotetext{I followed the introductory online Python course that is offered by Codeacademy, which is available at \href{https://www.codecademy.com/learn/learn-python}{https://www.codecademy.com/learn/learn-python}}

\leavevmode

\end{appendix}

\end{document}